\RequirePackage{fix-cm}
\documentclass[twocolumn,epj,nopacs]{svjour}
\smartqed  
\usepackage{url}
\usepackage{doi} 
\usepackage{hyperref}
\RequirePackage{graphicx}
\journalname{Prepared for EPJC}
\RequirePackage{xspace}
\RequirePackage{amssymb}
\RequirePackage{url}
\RequirePackage{siunitx}
\RequirePackage{lscape} 
\RequirePackage[inline]{enumitem} 
\RequirePackage{amssymb}
\RequirePackage{color}
\RequirePackage{lineno}
\usepackage{gensymb}

\newcommand{\cygno}{{\textsc{Cygno}}\xspace}

\newcommand{\Et}  {E$_{\mathrm{Transf}}$}
\newcommand{\Ed}  {E$_{\mathrm{Drift}}$}

\newcommand{\Vg}  {$V_{\mathrm{GEM}}$\xspace}
\newcommand{\Eg}  {$E_{\mathrm{GEM}}$\xspace}

\newcommand{\mn} {${^{55}\mathrm{Mn}}$\xspace}
\newcommand{\fe} {${^{55}\mathrm{Fe}}$\xspace}
\newcommand{\Isc} {$\SI{}{I_{SC}}$\xspace}

\newcommand{\zfe}  {$z_{^{55}\mathrm{Fe}}$\xspace}

\begin{document}

\title{Modeling the light response of an optically readout GEM based TPC for the CYGNO experiment}


\titlerunning{Modeling of the GEM-based TPC response}        
\author{
Fernando Domingues Amaro\inst{1} %
           \and
Rita Antonietti\inst{2,3} %
           \and
Elisabetta Baracchini \inst{4,5} %
           \and
Luigi  Benussi \inst{6} %
           \and
Stefano Bianco \inst{6} %
          \and
Roberto Campagnola \inst{6} %
          \and
Cesidio Capoccia \inst{6} %
          \and
Michele Caponero \inst{6,9} %
           \and
Gianluca Cavoto \inst{7,8} %
           \and
Igor Abritta Costa \inst{6} %
           \and
Antonio Croce \inst{6} %
           \and           
Emiliano Dan\'e \inst{6} %
           \and
Melba D'Astolfo \inst{4,5} %
           \and
Giorgio Dho \inst{6} %
           \and
Flaminia Di Giambattista \inst{4,5} %
           \and
Emanuele Di Marco \inst{7} %
          \and
Giulia D'Imperio \inst{7} %
           \and
Joaquim Marques Ferreira dos Santos \inst{1} %
           \and
Davide Fiorina \inst{4,5} %
           \and
Francesco Iacoangeli \inst{7} %
           \and
Zahoor Ul Islam \inst{4,5} %
           \and
Herman Pessoa Lima J\'unior \inst{4,5} %
           \and
Ernesto Kemp\inst{10}
       \and
Francesca Lewis\inst{8}
           \and
Giovanni Maccarrone \inst{6} %
           \and
Rui Daniel Passos Mano \inst{1} %
           \and
Robert Renz Marcelo Gregorio \inst{11} %
           \and
David Jos\'e Gaspar  Marques \inst{4,5} %
           \and
Luan Gomes Mattosinhos de Carvalho \inst{12} %
           \and
Giovanni Mazzitelli \inst{6} %
           \and
Alasdair Gregor McLean \inst{11} %
          \and
Pietro Meloni\inst{2,3} %
           \and
Andrea Messina \inst{7,8} %
           \and
Cristina Maria Bernardes Monteiro \inst{1} %
           \and
Rafael Antunes Nobrega \inst{12} %
           \and
Igor Fonseca Pains \inst{12} %
           \and
Matteo Pantalena\inst{8}
          \and
Emiliano Paoletti \inst{6} %
           \and
Luciano Passamonti \inst{6} %
           \and
Fabrizio Petrucci \inst{2,3} %
           \and
Stefano Piacentini \inst{4,5} %
           \and
Davide Piccolo \inst{6} %
           \and
Daniele Pierluigi \inst{6} %
           \and
Davide Pinci \inst{7}\thanks{\textit{corresponding author:} davide.pinci@roma1.infn.it}
           \and
Atul Prajapati \inst{4,5} %
           \and
Francesco Renga \inst{7} %
           \and
Rita Joana Cruz Roque \inst{1} %
           \and
Filippo Rosatelli \inst{6} %
           \and
Alessandro Russo \inst{6} %
           \and
Sabrina Salamino\inst{8}
          \and
Giovanna Saviano \inst{6,13} %
           \and
Federico Francesco Scamporlino\inst{8}
          \and
Angelo Serrecchia\inst{8}
          \and
Pedro Alberto Oliveira Costa Silva \inst{1}
           \and
Neil John Curwen Spooner \inst{11}
           \and
Roberto Tesauro \inst{6}
           \and
Sandro Tomassini \inst{6}
           \and
Samuele Torelli \inst{4,5}
           \and
Donatella Tozzi \inst{7,8}
}


\institute{LIBPhys, Department of Physics, University of Coimbra, 3004-516 Coimbra, Portugal; \label{1} 
           \and
Istituto Nazionale di Fisica Nucleare, Sezione di Roma TRE, 00146, Roma, Italy; \label{2}
            \and
 Dipartimento di Matematica e Fisica, Universit\`a Roma TRE, 00146, Roma, Italy; \label{3}
            \and
Gran Sasso Science Institute, 67100, L'Aquila, Italy; \label{4}
           \and
 Istituto Nazionale di Fisica Nucleare, Laboratori Nazionali del Gran Sasso, 67100, Assergi, Italy; \label{5}
            \and
 Istituto Nazionale di Fisica Nucleare, Laboratori Nazionali  di Frascati,  00044, Frascati, Italy; \label{6}
           \and
Istituto Nazionale di Fisica Nucleare, Sezione di Roma, 00185, Rome, Italy; \label{7}
           \and
  Dipartimento di Fisica, Sapienza Universit\`a di Roma, 00185, Roma, Italy; \label{8}
           \and
 ENEA Centro Ricerche Frascati, 00044, Frascati, Italy; \label{9}
           \and
Universidade Estadual de Campinas  - UNICAMP,  Campinas 13083-859, SP, Brazil; \label{10}
           \and
 Department of Physics and Astronomy, University of Sheffield, Sheffield, S3 7RH, UK; \label{11}
 \and
  Universidade Federal de Juiz de Fora, Faculdade de Engenharia, 36036-900, Juiz de Fora, MG, Brasil; \label{12}
           \and
 Dipartimento di Ingegneria Chimica, Materiali e Ambiente, Sapienza Universit\`a di Roma, 00185, Roma, Italy; \label{13}
}


\date{Received: 9 May 2025}



\abstract{
The use of gaseous Time Projection Chambers enables the detection and the detailed study of rare events due to particles interactions with the atoms of the gas with energy releases as low as a few keV. Due to this capability, these instruments are being developed for applications in the field of astroparticle physics, such as the study of dark matter and neutrinos. To acquire events occurring in the sensitive volume with a  high granularity,  the \cygno collaboration is developing a solution where the light generated during the avalanche processes occurring in a multiplication stage based on Gas Electron Multiplier (GEM) is read out by optical sensors with very high sensitivity and spatial resolution. 
To achieve a high light output, gas gain values of the order of $10^5\text{-}10^6$ are needed. 
In this working condition, a dependence of the detector response on the spatial density of the charge collected in the GEM holes has been observed, indicating a gain-reduction effect likely caused by space-charge buildup within the multiplication channels.
This paper presents data collected with a prototype featuring a sensitive volume of about two liters, together with a model developed by the collaboration to describe and predict the gain dependence on charge density. A comparison with experimental data shows that the model reproduces, with a percent-level precision, the gain behaviour over nearly one order of magnitude.}

\maketitle
%

\section{Introduction}

The \cygno  project \cite{bib:cygno} is aiming at the realisation of a cubic meter scale gaseous Time Projection Chamber (TPC) operating at atmospheric pressure for the search and study of rare events such as neutrino interactions or weakly-interacting massive particle (WIMP) scattering \cite{bib:history,bib:cygnus1,bib:cygnus2}. Gaseous TPCs are very suitable devices for these researches: they offer the possibility of instrumenting large sensitive volumes with a reduced number of readout channels compared to other approaches while retaining the possibility of having a complete reconstruction of the events within them, with high spatial and energy resolutions \cite{bib:TPCDM}.

Moreover, in a gas, a nuclear or electron recoil with an energy of a few \SI{}{keV}
would travel for hundreds to thousands 
of microns, leaving a trail of ionised atoms and free electrons that can be exploited to produce a detectable signal and allow for a three dimensional reconstruction of the particle direction.
A great deal of effort has been made in recent decades by various collaborations to develop devices capable of detecting low-energy signals while maintaining the ability to reconstruct their direction based, in particular, on the gaseous TPC technology  \cite{bib:Buckland,Battat:2014mka,Battat:2015rna,FENG201735,ALNER2005173,Ikeda:2020mvr,Battat:2014van}.
Although these efforts have yielded new and encouraging results, the main limitations in this development have been determined by the spatial resolution required to acquire the details of the tracks produced in the gas by particles of a few keV. Even when using low-pressure gas mixtures, an effective detector must be able to achieve a space resolution of a few millimetres over surfaces of several hundreds of square centimetres. Such high granularity is difficult to achieve with standard reading methods as requiring the use of tens or hundreds of thousands of independent channels.

\section{Optical readout}

One possible solution for obtaining the high granularity required is to switch to optical signal reading, collecting the light produced in the multiplication zones on appropriate optical sensors \cite{bib:Fraga,bib:opto1}.
In this respect, the \cygno project has conducted an R\&D program
to evaluate the feasibility and the performance of a gaseous TPC, with an amplification based on a Gas Electron Multiplier (GEM,  \cite{bib:sauli}) stage and an optical readout.

In particular, the idea is that the light produced by electroluminescence during the multiplication processes in the GEM channels with a readout system that
combines the high spatial resolution of the Active Pixel Sensors (APS, \cite{bib:aps}) and the time resolution of photomultipliers \cite{bib:jinst_orange1,bib:jinst_orange2}. 
The main advantage of this approach is related to the excellent performance that scientific CMOS-based (sCMOS) APS are able to provide:
\begin{itemize}
\item a high granularity with millions of independent readout pixels;
\item an average noise level and a sensitivity allowing the detection of individual photons with high efficiency;
\end{itemize}
Moreover, the optical coupling gives the possibility to keep the sensor out of the sensitive volume (no interference with HV operation and lower gas contamination)
and, by means of a suitable lens system, it is possible to acquire large surfaces with small sensors (reducing the setup complexity).

The geometrical optical acceptance $\epsilon_{\Omega}$ of a lens system, defined as the fraction of emitted photons reaching the sensor, depends on the lens aperture $a$\footnote{with the \textit{aperture}  of a lens (usually also denoted as $\#$) in optics is indicated the ratio between the focal length (FL) and the diaphragm diameter (D): $a$=FL/D} and the system magnification $\delta$ (the ratio between the captured area and the sensor area). It can be expressed as:

\begin{equation}
\label{eq:opto}
    \epsilon_{\Omega} = \frac{1}{(4(\delta+1)\cdot a)^2}
\end{equation}

 Typical setups with $\delta$  of the order of few tens and $a$ around 1 yield values of $\epsilon_{\Omega}=10^{-3}\text{-} 10^{-4}$.

\subsection{The He and CF$_4$ based gas mixture}

According to previous studies \cite{bib:Margato1,bib:brun}, 
electroluminescence spectra of CF$_4$-based mixtures show three main maxima: one around  a wavelength of \SI{250}{nm}, one around \SI{300}{nm} and one around  \SI{620}{nm}. 
Since this last wavelength matches the region of largest quantum efficiency of silicon-based light sensors, CF$_4$-based gas mixtures are commonly used for the optical readout. 

In particular, the \cygno collaboration has being using a CF$_4$ and helium mixture since 2015~\cite{bib:jinst_orange1}. Helium, thanks to its light nucleus, allows for large momentum transfers in collisions with particles of GeV-scale mass, making it especially suitable for studying low-mass WIMP interactions.

  The use of a proportion 60/40 between He and CF$_4$ makes it possible to operate under very stable conditions even with high gain in the amplification stages, leading to the production of up to 2-3 millions of secondary electrons per primary one \cite{bib:fe55}. 

  An average energy loss needed to produce an electron-ion pair ($w$ value) of \SI{35}{eV} is obtained by taking the weighted average of the He 
  and the CF$_4$ values
  \cite{Wvalue,Wvalue_thesis,WvalueHe,Wvalue_mixt}.
  
  For this particular gas mixture, 0.07 visible photons are expected to be produced per secondary electron in the avalanche \cite{bib:Fraga,bib:ieee_orange}.
Taking into account the values of $\epsilon_{\Omega}$ evaluated above, a large amount of secondary electrons is needed to ensure the collection of a few photons per electronvolt released into the gas, that would allow the detection of energy releases of the order of tens-hundreds of \SI{}{eV}.

On the other hand, the large amount of charges in the multiplication region is very likely responsible for the non-linear phenomena observed in the detector's response described in this paper.


\section{Experimental setup}

In recent years, the \cygno collaboration has developed several prototypes for R\&D studies on optical TPC technology, with sensitive volumes ranging from 0.1 liters (ORANGE, \cite{bib:nim_orange1,bib:nim_orange2} and MANGO \cite{bib:lumin1,bib:lumin2}), to 7 liters (LEMON \cite{coronello}) and then to 50 liters (LIME \cite{bib:lime_over}). A last prototype (called GIN) with a sensitive volume of approximately two liters, designed as a flexible and multi-purpose device for R\&D and tests, was recently realised by the INFN Laboratori Nazionali di Frascati (LNF).

\subsection{GIN detector}

\begin{figure*}[h]
    \centering
    \includegraphics[width=0.80\linewidth]{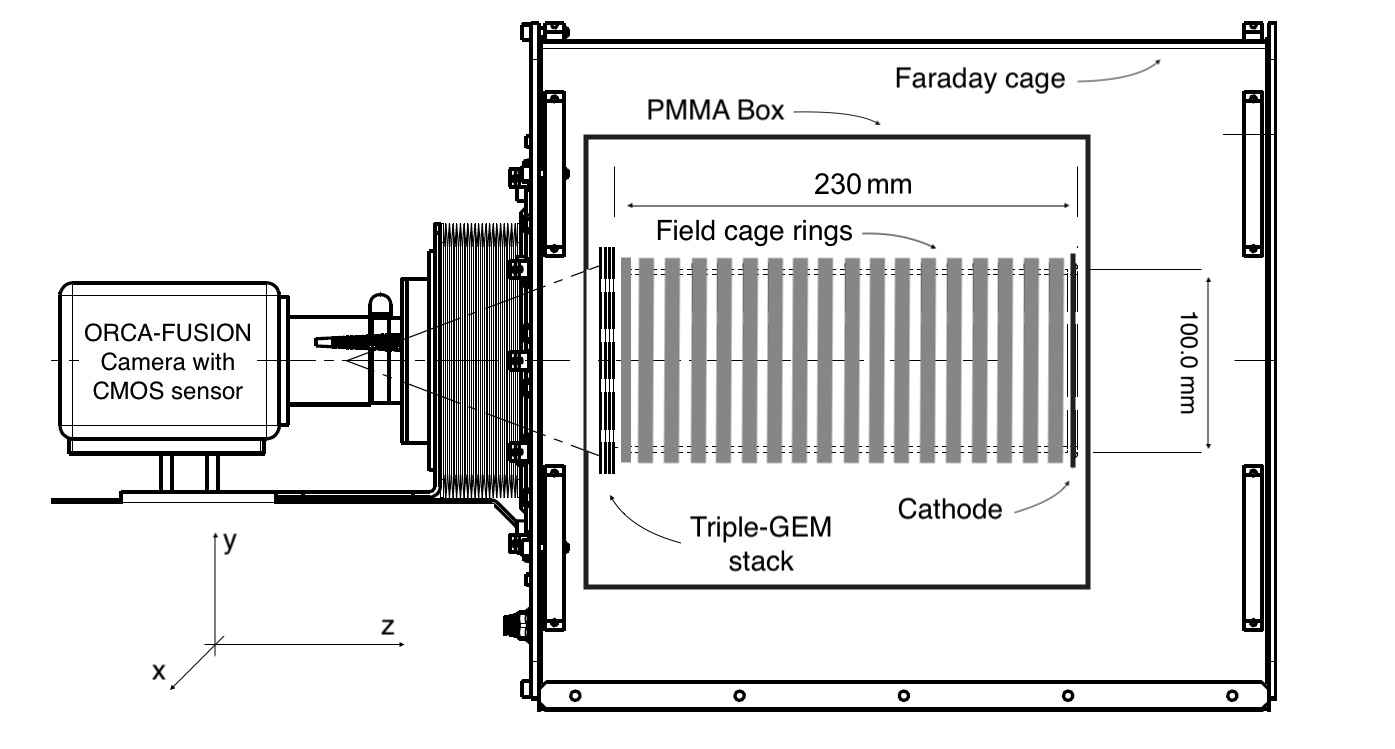}
    \caption{A lateral sketch of the GIN detector, showing the PMMA vessel and the field cage, the GEM plane and the photo-camera at the left and the cathode on the right.
    The used reference frame is shown at the bottom.}
    \label{fig:gin_draw}
\end{figure*}

The main components of the GIN prototype, depicted in Fig.~\ref{fig:gin_draw}, are described in the following subsections.

\subsubsection{Vessel, Field Cage and GEM stack}

The gas vessel is made of a transparent poly-methyl methacrylate (PMMA) parallelepipedal box. 

Within the acrylic vessel, a series of seventeen $10 \times 10$ \SI{}{cm^2}  squared copper rings electrically connected in series through resistors and  maintained at progressively increasing potential values act as electrodes of an electric-field cage (FC). The FC is closed on one side by a $10 \times 10$~cm$^2$ copper cathode, which defines the lowest potential, and on the other side by a structure of 
three $10 \times 10$~\SI{}{cm^2} GEMs.
These have bi-conical multiplication channels with an internal diameter of \SI{50}{\mu m}, an external one of \SI{70}{\mu m} and pitch of \SI{140}{\mu m}. The three GEMs are placed \SI{5}{mm} apart from the first field cage coil and \SI{2}{mm} apart from each other. The gaps between the three GEMs are usually referred to as \textit{transfer gaps}.
This configuration generates within the field cage a uniform electric field (called \textit{drift field} \Ed), oriented orthogonally to the cathode and GEM planes, that guides the ionization electrons produced by charged particles traversing the gas towards the anodic GEMs stack.
The drift field region spans \SI{230}{mm} in length making the sensitive gas volume of the GIN prototype to be \SI{2.2}{liters}.
The bottom plane of GEM$_3$ is kept to ground potential and all electrons extracted from last GEM are collected in it.

Electrical potentials at the various electrodes of the detector are supplied by two generators:
\begin{itemize}
\item cathode: an ISEG \textit{HPn 500}\footnote{For more details, please visit https://iseg-hv.com/en/home} provides up to 50 kV and 7 mA with negative polarity and ripple $<0.2\%$;
\item GEM electrodes: CAEN \textit{A1515TG}\footnote{For more details, please visit https://www.caen.it/families/universal-multichannel-system/} board with individual floating channels supplies the voltages (up to 1 kV with 20 mV precision);
\end{itemize} 

By means of these two power suppliers, the following electrostatic configuration is setup in GIN:
\begin{itemize}
\item  a constant \Ed with a typical value of \SI{1.0}{kV/cm} in the sensitive volume;
\item  a constant \textit{transfer field} in the transfer gaps between the GEMs (\Et~=~\SI{2.5}{kV/cm});
\item the voltage difference across the two sides of each GEM (\Vg) that was changed accordingly to the different test needs.
\end{itemize}

\subsubsection{Light sensors}
\label{sec:light_sen}

The light produced in the multiplication channels of the last GEM is transmitted out of the vessel by a \SI{200}{\mu m} thick PET window.
A black optical bellow is mounted on the external side of the window and allows for a safe transmission of the light signal to the optical sensors: 

\begin{itemize}
    \item an ORCA-Fusion\footnote{For more details please visit https://www.hamamatsu.com/eu/en/product/cameras/cmos-cameras/C14440-20UP.html} scientific CMOS camera with $2304 \times 2304$ pixels of $6.5 \times 6.5$ \SI{}{\mu m^2} 
    able to detect photons over the whole visible spectrum (while it is blind below \SI{320}{nm}) with quantum efficiency showing a peak of 80\% at \SI{500}{nm} and an average $\bar{QE}$ value of about 75\% for region around \SI{620}{nm} of the He/CF$_4$ emission spectrum. Placed at a distance of \SI{22.3}{cm} from the last GEM, and equipped with a Schneider Xenon lens (\SI{25.6}{mm} focal length $f$, an aperture $a$ of 0.95 and an almost flat transmittance $T_L$ of about 90\% in in the \SI{500}{nm} - \SI{700}{nm} range), the camera captures an $11.52 \times 11.52$ \SI{}{cm^2} image with a
    the geometrical acceptance $\epsilon_{\Omega}$ of $9.2 \times 10^{-4}$. In this configuration, each pixel can frame an area of $50 \times 50$ \SI{}{\mu m^2}.
    
    \item two Hamamatsu R1894 photomultipliers, with a \SI{10}{mm} diameter, with a spectral response from \SI{300}{nm} to \SI{850}{nm} and a QE of about 1\% at \SI{620}{nm}.
    \end{itemize}

\subsubsection{Gas supply system}

The gas vessel is continuously flushed at a rate of \SI{150}{cc/min} with a  mixture, obtained from bottles of  pure gases: He with a rate of \SI{90}{cc/min} and CF$_4$ with a rate of \SI{60}{cc/min}. No recirculation of the gas is foreseen and the  output gas  is sent to an exhaust line connected to the external environment via a water-filled bubbler ensuring an over-pressure of approximately \SI{3}{mbar}, relative to the external atmospheric pressure.

\subsubsection{Source window}

\begin{figure}[h]
    \centering
    \includegraphics[width=0.95\linewidth]{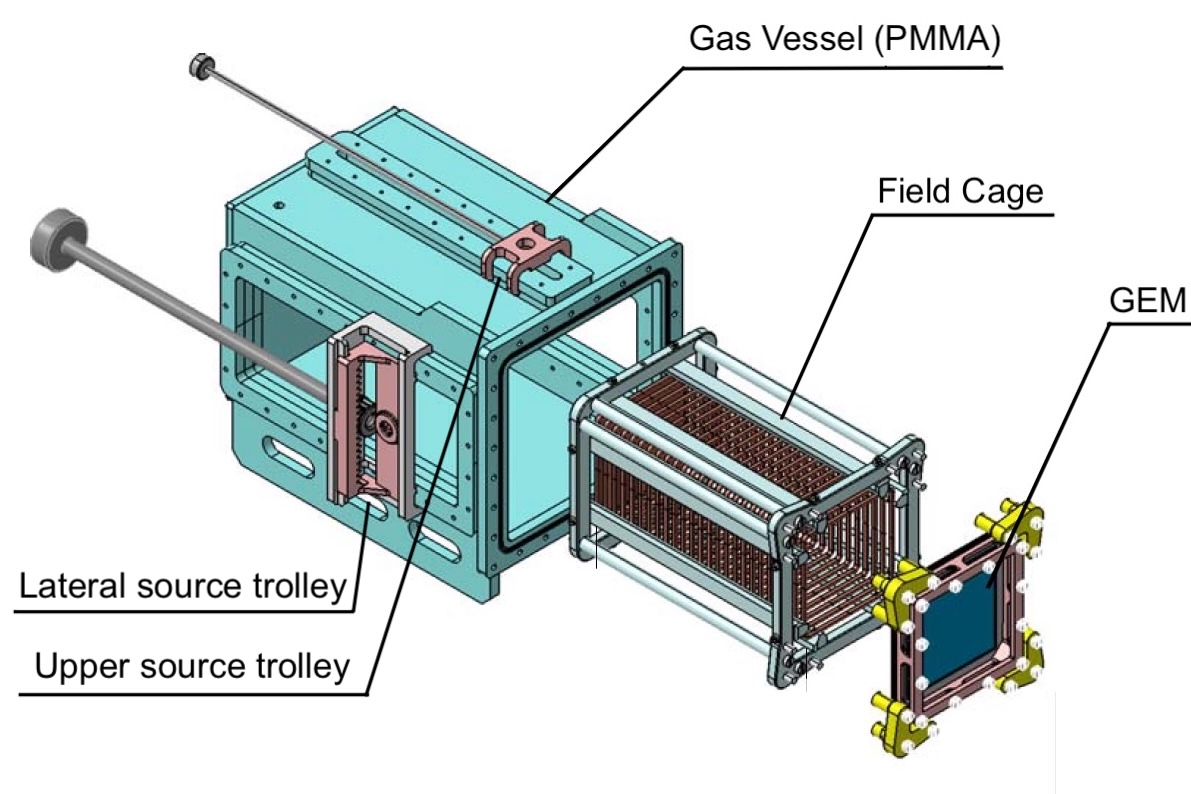}
    \caption{Exploded view of the GIN detector, showing in particular the two trolleys and corresponding rods for the movement on two thin windows (see text), foreseen for tests with radioactive sources. In this work, only the upper one was present and used for the \fe source.}
    \label{fig:gin_exploded}
\end{figure}

The upper surface of the vessel features a thin window, \SI{2}{cm} wide and \SI{23}{cm} long perpendicular to the GEM plane, sealed by a \SI{125}{\mu m} thick layer of ethylene-tetrafluoroethylene (ETFE) (see Fig. \ref{fig:gin_exploded}). This window permits the entry of a reasonable fraction of low-energy photons (order of 10\% for few keV) into the gas volume from external radioactive sources utilized for calibration purposes.

A manually controlled trolley is mounted above this window and is capable of moving back and forth on a predefined rail along the $z$ axis. It serves as a holder for the radioactive source and enables its movement, positioned \SI{8}{cm} above the sensitive volume, ranging from $z$=\SI{1}{cm} to $z$=\SI{23}{cm} from the GEMs. 

\subsubsection{Faraday cage}

The light tightness and electrical shield of the detector is ensured by an external Faraday cage made of a \SI{3}{mm} thick aluminum metal box suitably equipped with feed-through connections for the high voltages required for the GEMs, cathode and PMT and for the gas piping.
A rod, free to enter through the opposite side to the camera, allows the movement of the source trolley.

\section{Data taking}

To study the behaviour of the detector light response, the signals produced by the interaction of X-rays in the gas volume were acquired and analysed.

\subsection{Data acquisition system}
\label{sect:daq}
The data acquisition is implemented by using an integrated system within the Midas framework\footnote{for more details please visit  https://daq00.triumf.ca/MidasWiki/index.php/Main\_Page}. 
PMT analogue signals are routed to a discriminator and then to a logic module, which generates a trigger signal based on the coincidence of the signals above threshold from the two PMTs. A dedicated data acquisition PC is connected via two USB 3.0 ports to both the camera and a VME crate containing I/O register modules for triggering and control functions. Although the DAQ system is designed to accommodate digitizers for acquiring PMT signal waveforms, this article focuses exclusively on the analysis of 2D image obtained by the camera.

\subsection{Detector Operation}
\label{sect:operation}
The measurements reported in this paper were carried out at the INFN-LNF laboratories. The detector was operated inside an experimental hall where the temperature varied in a range  between \SI{295}{K} and \SI{300}{K} and the atmospheric pressure between \SI{970}{mbar} and \SI{1000}{mbar} for the entire duration of the measurements. The typical working conditions of the detector are reported in Table \ref{tab:parameter}.

\begin{table}[h]
\centering
\caption{Summary of the typical operating condition of the GIN detector during the data takings.\label{tab:parameter}}
\vspace{1mm}
\begin{tabular}{l c}
\hline
Parameter      & Typical value \\ \hline \hline
Drift Field    & 1.0 kV/cm      \\ \hline
GEM Voltage    & 420 V - 440 V         \\ \hline
Transfer Field & 2.5 kV/cm     \\ \hline
Gas Flow       & 9.0 l/h       \\ \hline
PMT Threshold  & 15 mV         \\ \hline
\end{tabular}
\end{table}

\subsection{Data acquisition strategy}

The camera supports adjustable exposure time. Following an optimization designed to balance the average number of events per image, keeping it low enough to minimize the probability of overlapping events, yet high enough to avoid an excessive number of empty frames, an exposure of \SI{150}{ms} was set.

\begin{figure}[h]
    \centering
    \includegraphics[width=0.9\linewidth]{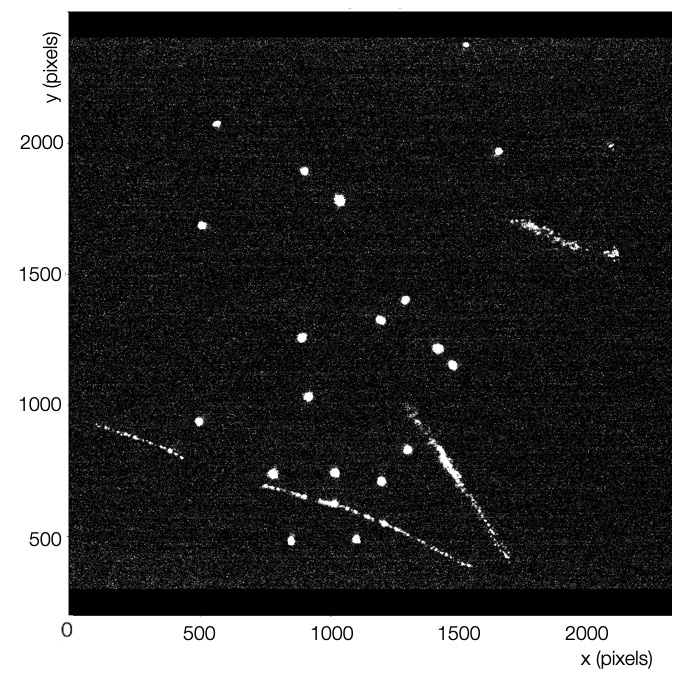}
    \caption{Example of a typical image collected by the camera during the tests (a pixel is equivalent to about $50 \times 50$ \SI{}{\mu m^2}).}
    \label{fig:img_orig}
\end{figure}

Figure \ref{fig:img_orig} shows an example of a typical image taken during the tests. As described below, the small spots are expected to be due to low energy photons interaction in the gas, while long tracks are expected to be due to high energy electrons produced by natural radioactivity or cosmic rays muons.

The data acquisition is subdivided in \textit{runs}. Each run corresponds to a sample of 400 pictures taken by keeping the detector setup (high voltage values, gas flow, source position) in a stable configuration.

\subsection{The \fe source}
\label{sec:fe}
To study the detector response to specific energy deposits in the gas, an \fe source with a current activity of about \SI{100}{MBq} was employed. \fe decays to \mn, producing primarily photons with an energy of \SI{5.9}{keV} (K$_{\alpha}$), with a small fraction of photons at \SI{6.4}{keV} (K$_{\beta}$).
The source is made by a copper cylinder with \SI{15}{mm} diameter, \SI{15}{mm} height and has an active deposit of \SI{10}{mm} diameter on the bottom face, enclosed with a thin beryllium window.
These photons have a mean free path of approximately \SI{22}{cm} in the GIN gas mixture \cite{bib:lime_over} and interact with the gas molecules mainly through the photoelectric effect, producing electron recoils with similar energy. 
Given the average cost of about \SI{35}{eV} to produce an electron-ion pair, approximately 168 primary electrons (n$_e$) are produced in a range of the order of \SI{0.5}{mm}, which represents the expected mean range  for electrons of around \SI{6}{keV} in a He/CF$_4$ (60/40) mixture at atmospheric pressure.

\section{The detector optical response}
\label{sec:optical}

Under the effect of the drift field, the electrons produced by the absorption of the \fe photons in the gas start drifting towards the first GEM, where they are collected within the multiplication channels and initiate the avalanche process.

 During their drift, electrons are diffused by the scattering with gas molecules. As a result, the electron cluster will occupy a region  that depends on the distance of the interaction point to the GEMs. Their $x$-$y$ space distribution can be described with a 2-dimensions Gaussian profile having the same standard deviations on $x$ and $y$ ($\sigma_x = \sigma_y = \sigma$). The value of $\sigma$ depends on the drift distance $z$ through the equation: 
 
\begin{equation}
    \label{eq:sigma}
    \sigma = \sqrt{\sigma_0^2 + D^2_T \cdot z}
\end{equation}
where $\sigma_0$ is the minimum spread that has a contribution due the photoelectron ionization cloud (evaluated with GEANT4~\cite{bib:geant4} to be of the order of \SI{0.1}{mm^2}) and one  due to the diffusion happening in the GEM stack.
$D_T$ is the transverse diffusion coefficient, which depends on the gas mixture and the drift field \cite{bib:yellow}.

 
As already mentioned, on average, for every 100 secondary electrons produced in these multiplication processes, approximately 7 photons are also produced isotropically by the gas. According to the geometrical acceptance ($\epsilon_{\Omega}$), a fraction of the photons produced at the exit of the third GEM, n$_p$, are collected on the objective lens, contributing to the image formation on the optical sensor. Thus, given a certain energy release $E$, a proportionality is expected between it, the number of primary electrons n$_{e}$=$E/w$ and the number of photo-electrons (n$_{p.e.}$) detected by the optical sensor:

\begin{equation}
    \label{eq:fiorina1}
    \mathrm{n}_{p.e.} = \overline{QE} \cdot T_L \cdot T_{w} \cdot \epsilon_{\Omega} \cdot 0.07 \cdot G_{tot} \cdot n_{e}
\end{equation}
where $G_{tot}$ is the total effective electron gain of the triple-GEM stack and $T_{w}$ is the transparency of the vessel window measured to be in the range 0.94\%-0.96\% for wavelengths larger than \SI{400}{nm}.

For each \fe photon interacting with the gas in the sensitive area of the detector, we expect to observe an image on the sensor containing n$_{p.e.}$ photo-electrons distributed over a circular area, with a density that can be described by a 2D Gaussian profile having a $\sigma$ dependent on $z$.

\begin{figure}[h!]
    \centering
    \includegraphics[width=0.48\linewidth]{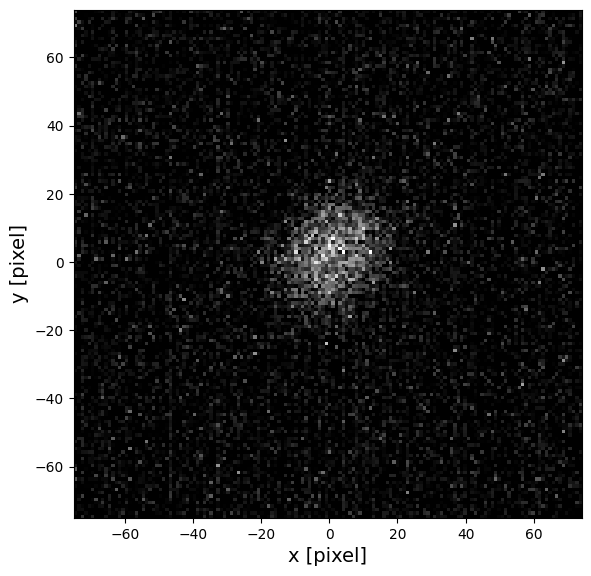} 
    \includegraphics[width=0.45\linewidth]{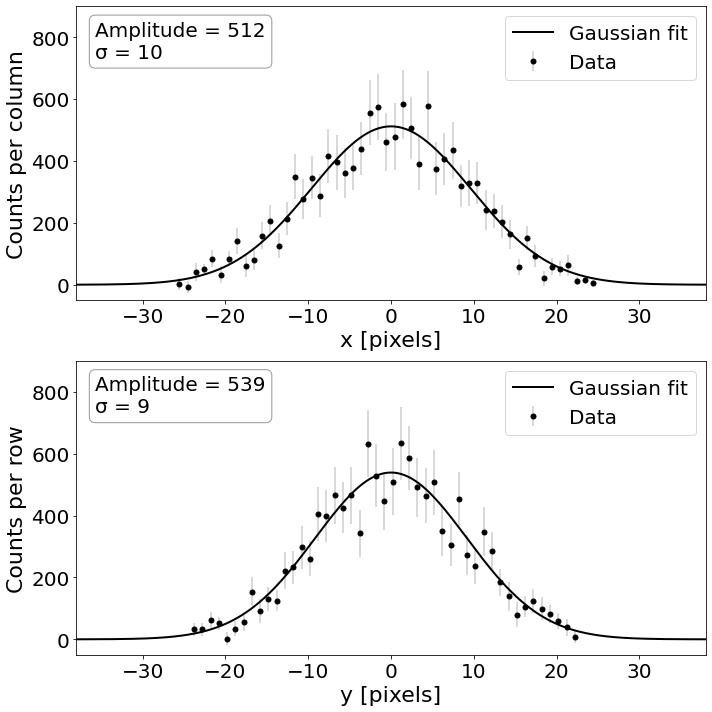} \\
    \includegraphics[width=0.48\linewidth]{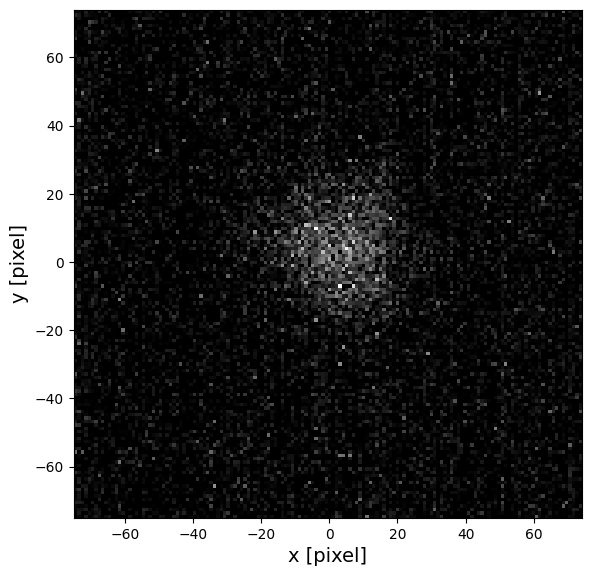} 
    \includegraphics[width=0.45\linewidth]{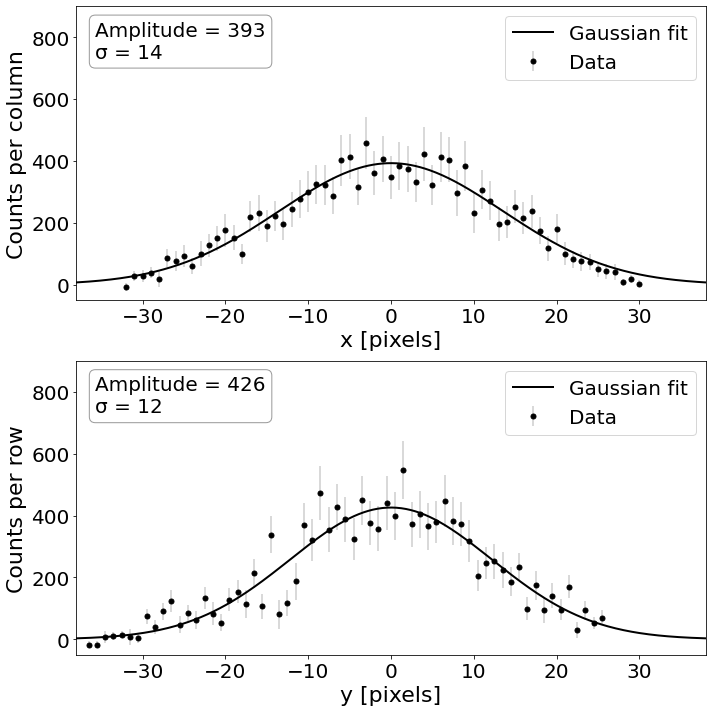} \\
    \caption{Left: two examples of a single \fe-induced cluster acquired by the camera with source placed at two different distances ($z=$\SI{6}{cm} and $z=$\SI{22}{cm}) from the GEM plane. 
     Right: the corresponding light profiles along $x$ and $y$ directions of the clusters, with superimposed Gaussian fits (a pixel is equivalent to about $50 \times 50$ \SI{}{\mu m^2}).}
    \label{fig:spot-profile}
\end{figure}

Figure \ref{fig:spot-profile} shows on the left two typical examples of light spot as recorded by the optical sensor and produced by the photon interaction in the gas with the \fe source at about \SI{4}{cm} (top) and at about \SI{22}{cm} (bottom) from the GEM plane. 
For both cases, the corresponding light profiles along the $x$ and $y$ directions are shown on the right.

The effect of the electron diffusion in the gas is clearly visible: interactions happening farther from the GEMs produce wider spots with a lower maximum number of counts in the central pixels.

\section{Cluster reconstruction and analysis}
\label{sec:reco}
As shown in Fig.~\ref{fig:img_orig}, in general, images with an exposure time of \SI{150}{ms} display various signals. Together with the $^{55}$Fe-induced clusters with their characteristic round-shape energy deposit, there can be signals generated by interactions in the gas involving gamma rays, X-rays, electrons, or alpha particles due to natural radioactivity, as well as muons from cosmic rays.
 While muons typically create straight and slim tracks with lengths comparable to the transverse size of the detector, the alphas produce short and very bright tracks. The 
 interactions due to natural radioactivity, given their lower energy, are mostly associated with irregular clusters. 
In order to select interactions produced by \fe, an algorithm developed by the collaboration \cite{bib:clus} was employed, capable of identifying groups of pixels hit by photons (clusters) in the images recorded on the sensor, as shown in Fig.~\ref{fig:spot-profile}.

\begin{figure}[h!]
    \centering
    \includegraphics[width=0.9\linewidth]{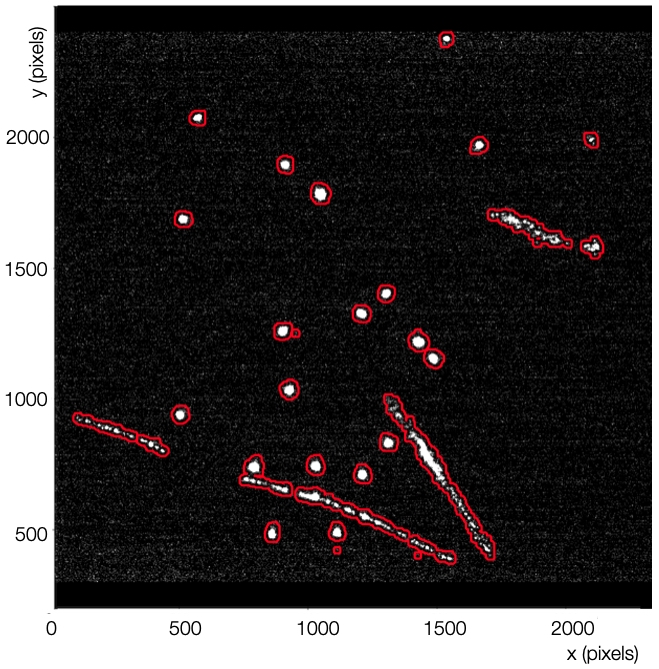}
    \caption{Example of an image from the camera with superimposed in red the clusters recognised by the reconstruction algorithm (a pixel is equivalent to about $50 \times 50$ \SI{}{\mu m^2}). }
    \label{fig:reco}
\end{figure}

After removing all pixels whose content is deemed compatible with the sensor's electronic noise (zero-suppression), the remaining pixels are processed using this algorithm. The algorithm is based on an upgraded version of DBSCAN \cite{dbscan}, which, in addition to considering the spatial density of pixels, weights them on the basis of the amount of collected hits \cite{iDBSCAN}. All pixels identified as belonging to the same spot form a cluster. Figure \ref{fig:reco} shows an example of an image containing several light spots with the corresponding reconstructed cluster borders superimposed in red.

For all clusters identified as such by the reconstruction algorithm, various characteristics are  calculated. Those relevant to the work described in this paper are listed below:  

\begin{itemize}  
\item \textbf{Integral (\Isc)}: the sum of all counts collected by all pixels belonging to the cluster. This variable essentially represents the total light collected in the cluster.  
\item \textbf{Length} and \textbf{Width}: these represent the lengths of the major and minor axes, extracted via a Principal Component Analysis (PCA). 
\item \textbf{Slimness (\(\xi\))} evaluated from the ratio (width/length), is related to the cluster shape: the closer it is to one, the more circular the cluster is.  
\item \textbf{Sigma}: the $\sigma$ of the Gaussian fit performed on the light profile transverse to the major axis. 

\end{itemize}  

\begin{figure}[h!]
    \centering
    \includegraphics[width=0.95\linewidth]{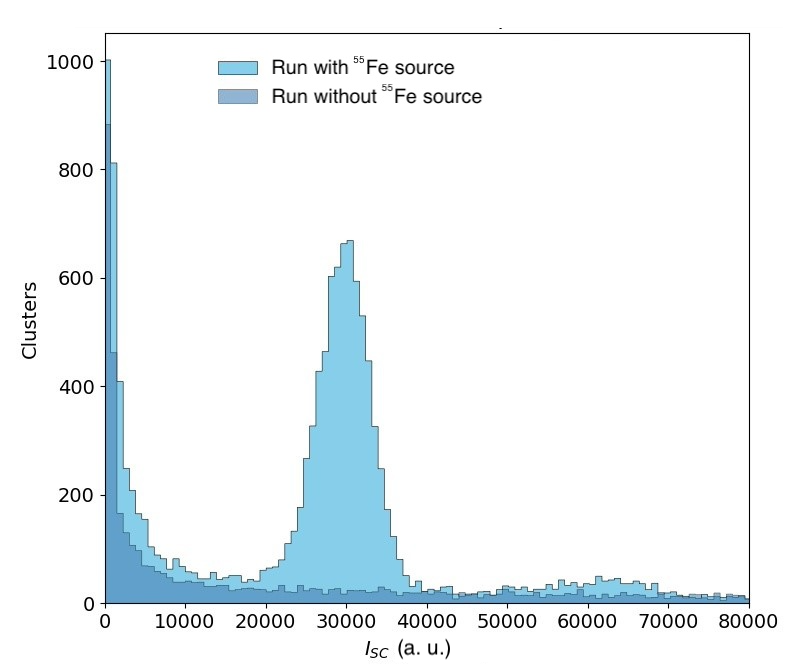}
    \caption{Distribution of the cluster light integral $I_{SC}$ for data runs with (light blue) and without (royal blue) the \fe source.}
    \label{fig:cuts}
\end{figure}

In Fig.~\ref{fig:cuts}, an example of the distributions of the cluster light integral $I_{SC}$ for runs with and without the \fe source is shown. The clear peak appearing at around 30,000 counts is the one related to absorption of the 5.9~keV. A second bump is visible around 65,000 counts  due to two or more rounded \fe clusters that are close to each other and are merged by the reconstruction algorithm, resulting in a single cluster.

\section{Signal selection}
\label{sec:sel}

By using the reconstructed variables for each cluster, the selection of spots that can be associated with photon interactions from the \fe source is carried out. 
Based on studies and optimisation performed in previous studies \cite{bib:fe55,Costa:2019tnu}, the following conditions are applied:

\begin{itemize}

\item In order to reject fake clusters resulting from electronic noise overfluctuations in the light sensor, a lower threshold of \Isc larger than 10,000 units is imposed. Simultaneously, to reduce the selection of merged-clusters, an upper limit of \Isc lower than 60,000 is required.

\item A maximum length of \SI{7.5}{mm} is required. Since the spots produced by the source are expected to have dimensions of a few \(\mathrm{mm}^2\), tracks longer than the optimised value are excluded to rejection most cosmic muon tracks while still accepting tracks from \fe photon interactions, whose width is larger than the expected range in the gas because of the diffusion effect.

\item The parameter \(\xi\) is considered, where values closer to 0 correspond to a more elongated cluster, such as the elongated tracks produced by muons, while values near to 1 represent more round clusters. As found in previous studies \cite{coronello}, the \(\xi\) value for clusters induced by the source peaks at approximately \(\xi = 0.9\), and the range \(0.7 < \xi < 1\) contains the majority of \fe source clusters.

\end{itemize}

\section{Position of the interaction points}
\label{sec:pos}
\begin{figure}[h]
    \includegraphics[width=0.95\linewidth]{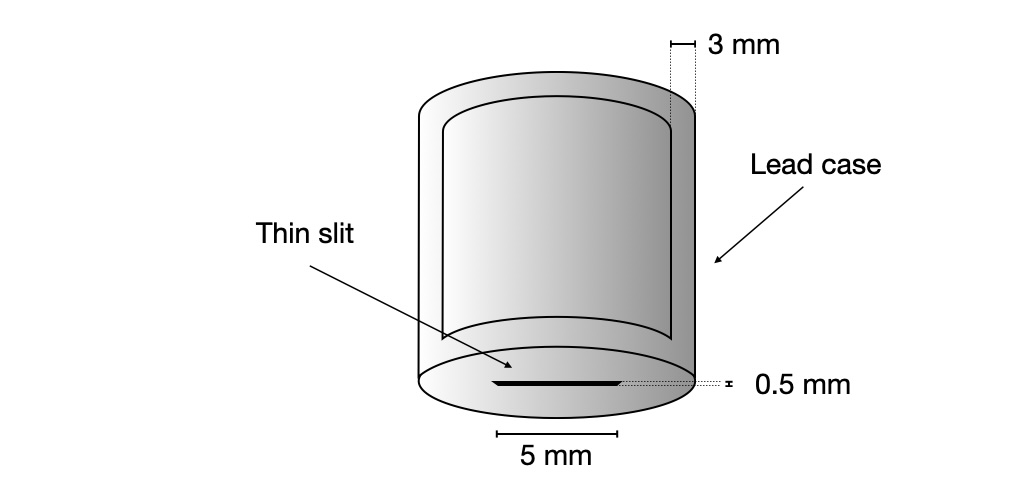}
    \caption{Sketch (not to scale) of the lead case used collimate the photon flux exiting from the \fe source.}
    \label{fig:coll}
\end{figure}

For the tests described in this article, a special lead case was used  as shown in Fig.~\ref{fig:coll}. 
The case features on the bottom face a rectangular slit measuring $0.5 \, \mathrm{mm} \times 5 \, \mathrm{mm}$. This design allows for relatively uniform illumination of the sensitive volume in the plane parallel to the slit’s long side, while simultaneously selecting a well-defined region in the orthogonal direction.
By means of an internal spacer, the source cylinder (see Sect.~\ref{sec:fe} for details) is held at a height of \SI{2}{cm} above the slit.

\begin{figure}[h]
    \includegraphics[width=1.\linewidth]{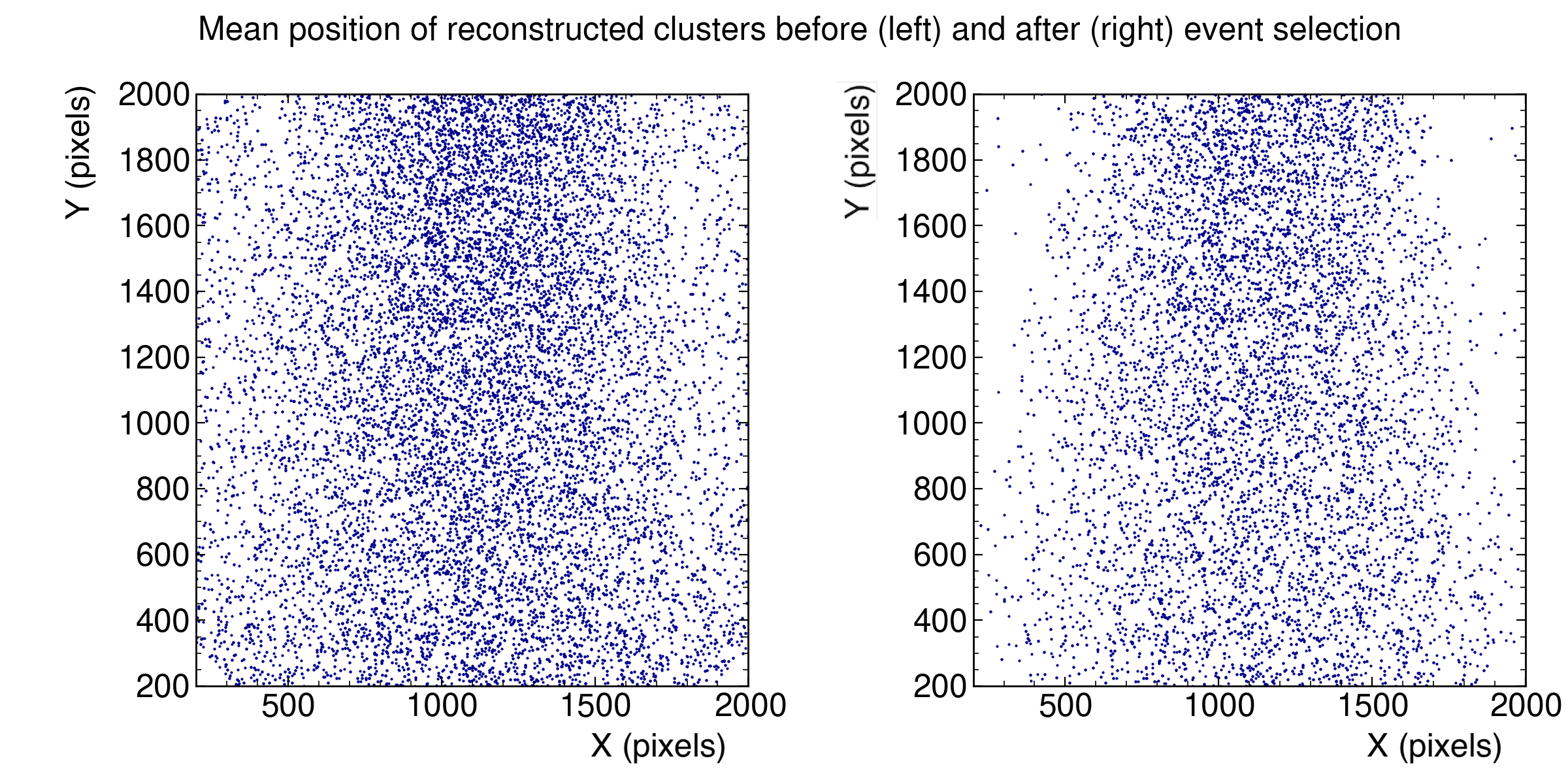}
    \caption{Maps of the positions of all the reconstructed clusters (left) and of the ones expected to be due to the interaction of the \fe photons (right) with the collimator slit parallel to the GEM plane (a pixel is equivalent to about $50 \times 50$ \SI{}{\mu m^2}).}
    \label{fig:cone-xy}
\end{figure}

\begin{figure}[h]
    \centering
    \includegraphics[width=0.90\linewidth]{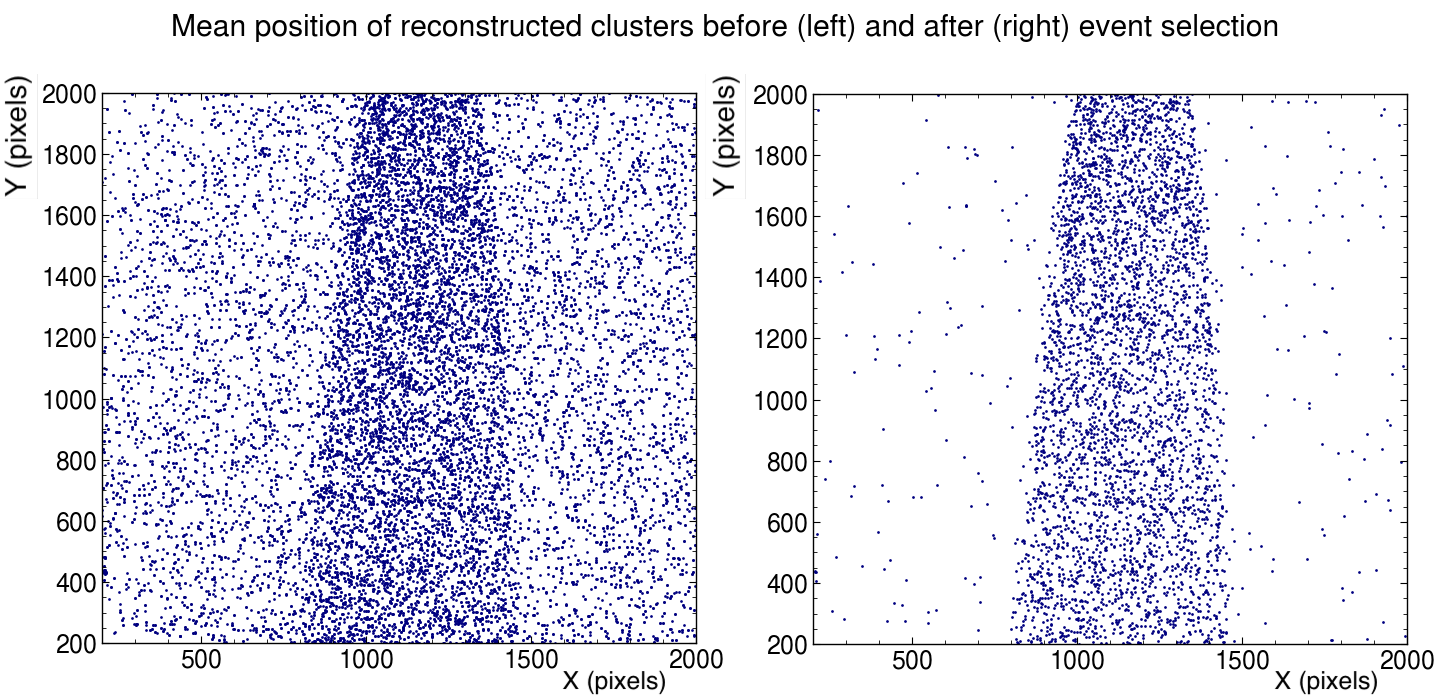}
    \caption{Maps of the positions of all the reconstructed clusters (left) and of the ones expected to be due to the interaction of the \fe photons (right) with the collimator slit orthogonal to the GEM plane (a pixel is equivalent to about $50 \times 50$ \SI{}{\mu m^2}).}
    \label{fig:cone-z}
\end{figure}

Figures \ref{fig:cone-xy} and \ref{fig:cone-z} illustrate, the reconstructed positions on the $x$-$y$ plane for all events (left panels) and for those selected as due to the \fe\ source (right panels), with the collimator oriented in two configurations:

\begin{itemize}
    \item \textbf{Figure \ref{fig:cone-xy}}: the collimator is positioned with the slit’s long side parallel to the GEMs. In this configuration, the interactions are distributed fairly uniformly, spanning $5 \, \mathrm{cm}$ upwards and $8 \, \mathrm{cm}$ downwards, covering approximately $65\%$ of the total area.
    \item \textbf{Figure \ref{fig:cone-z}}: the collimator is rotated such that the slit’s long side is orthogonal to the GEMs. Here, the effect of the shorter dimension is evident, with interactions concentrated within a region measuring $1.7 \, \mathrm{cm}$ upwards and $3.4 \, \mathrm{cm}$ downwards. 
\end{itemize}

In both configurations, the selection process effectively rejects the majority of events that are not expected to originate from the \fe\ source which are expected to be uniformly distributed in the $x$-$y$ plane, as observed in runs without the source.

An average RMS of the z-position distribution of the interaction points of about \SI{0.5}{cm} was then evaluated from data in Fig.~\ref{fig:cone-z} when the collimator slit is parallel to the GEM plane.

\section{The detector response as a function of the interaction position}

Using the \fe source, the detector response as a function of the ionization point distance from the GEM plane has been studied.
For this study, data-runs were collected both in the absence of the \fe source and in its presence, placing the source at ten different distances from the GEM plane (\zfe), spaced \SI{2}{cm} apart, ranging from about \SI{4}{cm} to about \SI{22}{cm}.
Data were taken with the slit orientation parallel to the GEM plane. As evaluated in previous section, this will produce a spread in the \zfe with an average RMS of about \SI{0.5}{cm}.

\subsection{Spot size and electron diffusion}

As discussed in Sect.~\ref{sec:optical} and shown in Eq.~\ref{eq:sigma}, primary electron diffuse during their drift along $z$ towards the GEMs. They are therefore expected to spread over a region on the GEM plane that increases with the drift distance.

For each run, after applying the event selection described in Sect.~\ref{sec:sel},  for all clusters attributed to \fe, the quantities introduced in Sect.~\ref{sec:reco} are then evaluated. 

\begin{figure}[h!]
    \centering
    \includegraphics[width=0.49\linewidth]{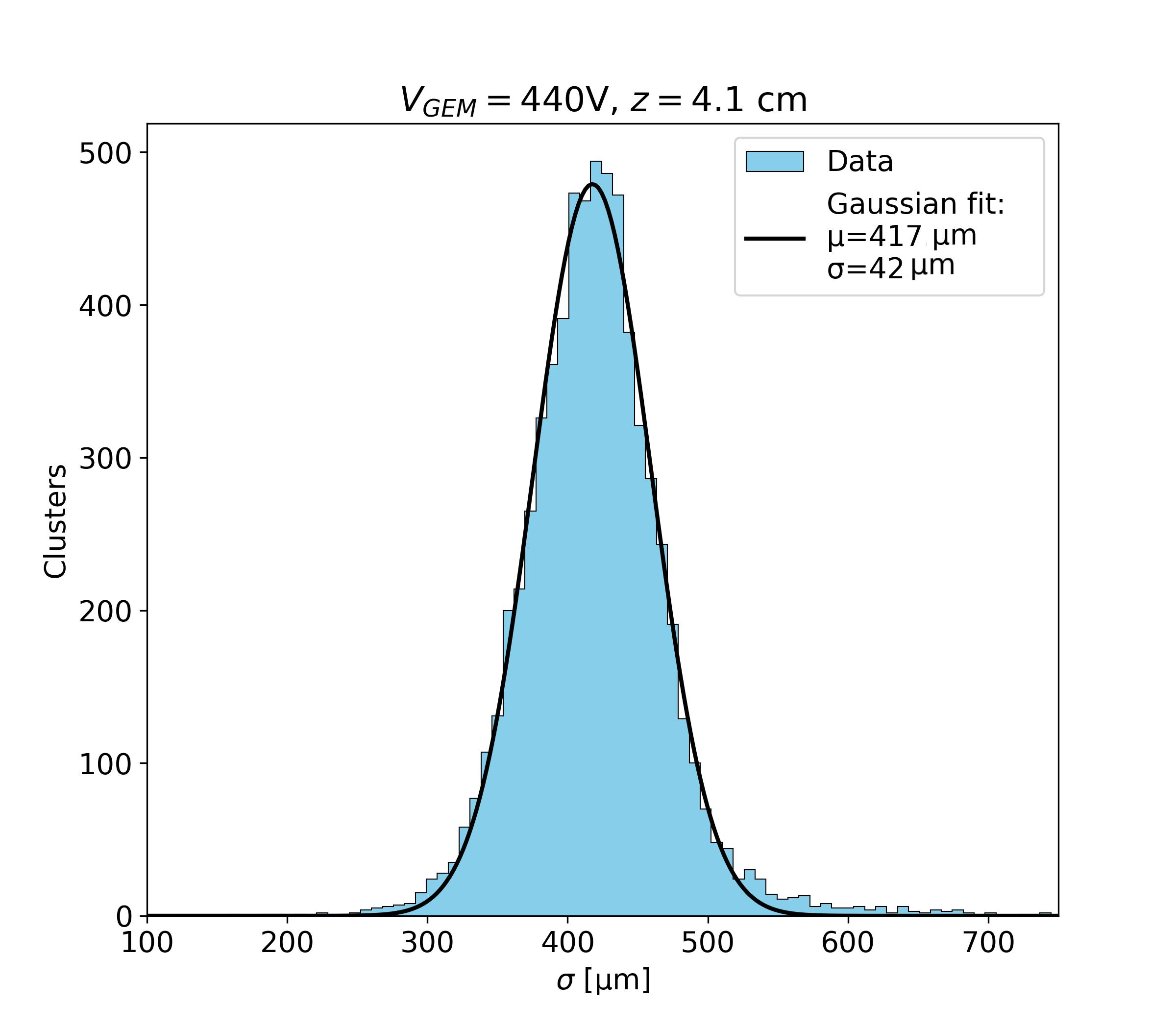}
    \includegraphics[width=0.49\linewidth]{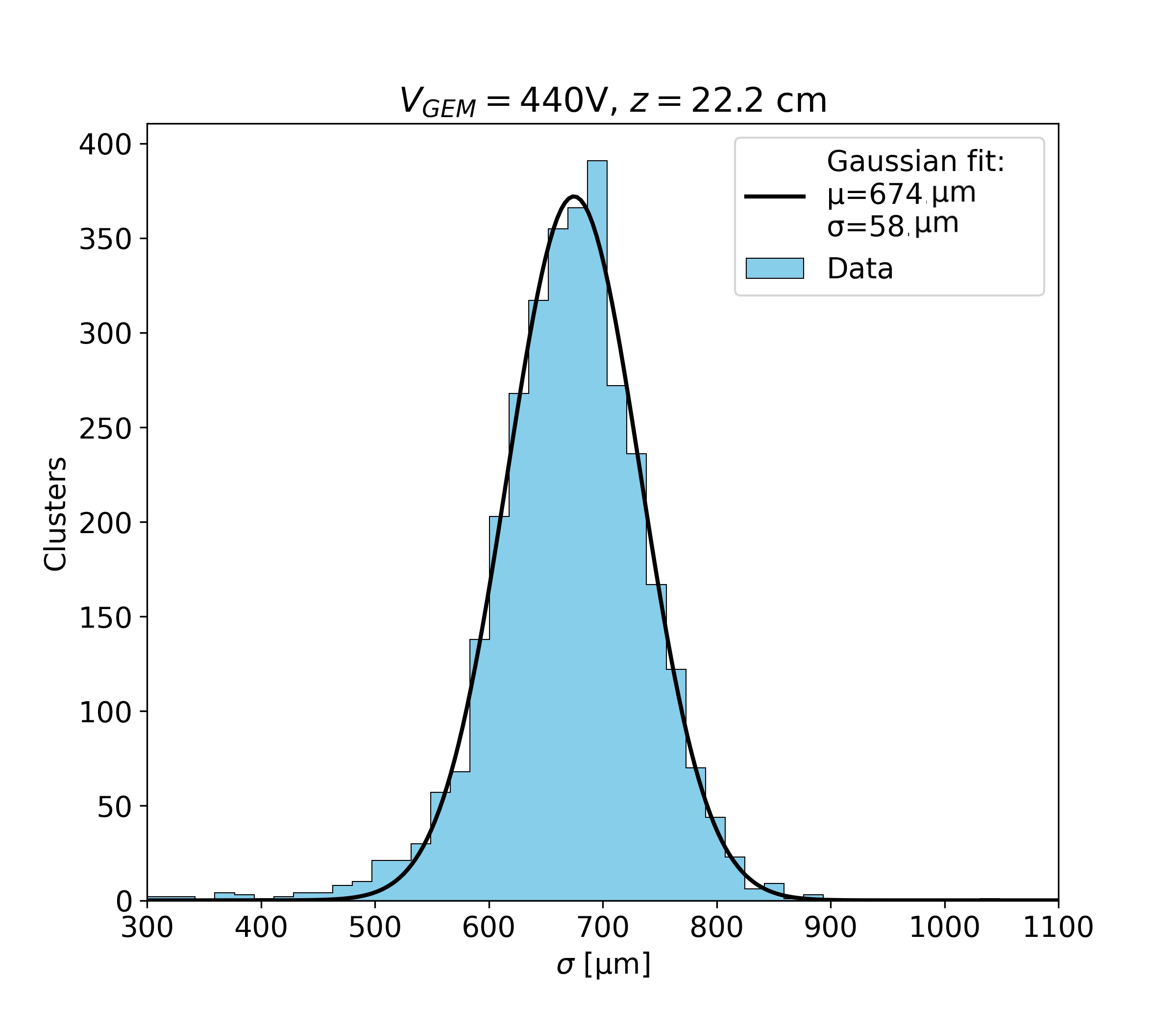}
    \caption{Distributions of the $\sigma$ values evaluated for the clusters attributed to the \fe interactions for the source positions \zfe=\SI{2.1}{cm} (left) and \zfe=\SI{22.2}{cm}  (right) with a superimposed Gaussian fit.}
    \label{fig:hsigma}
\end{figure}

In Fig.~\ref{fig:hsigma} two examples of distributions of $\sigma$ values   are shown  for interactions occurred nearer (left) or farther (right) to the GEM plane. It can be clearly seen that due to the effect of the diffusion the mean value increases for large distances.

\begin{figure}[h!]
    \centering
    \includegraphics[width=0.90\linewidth]{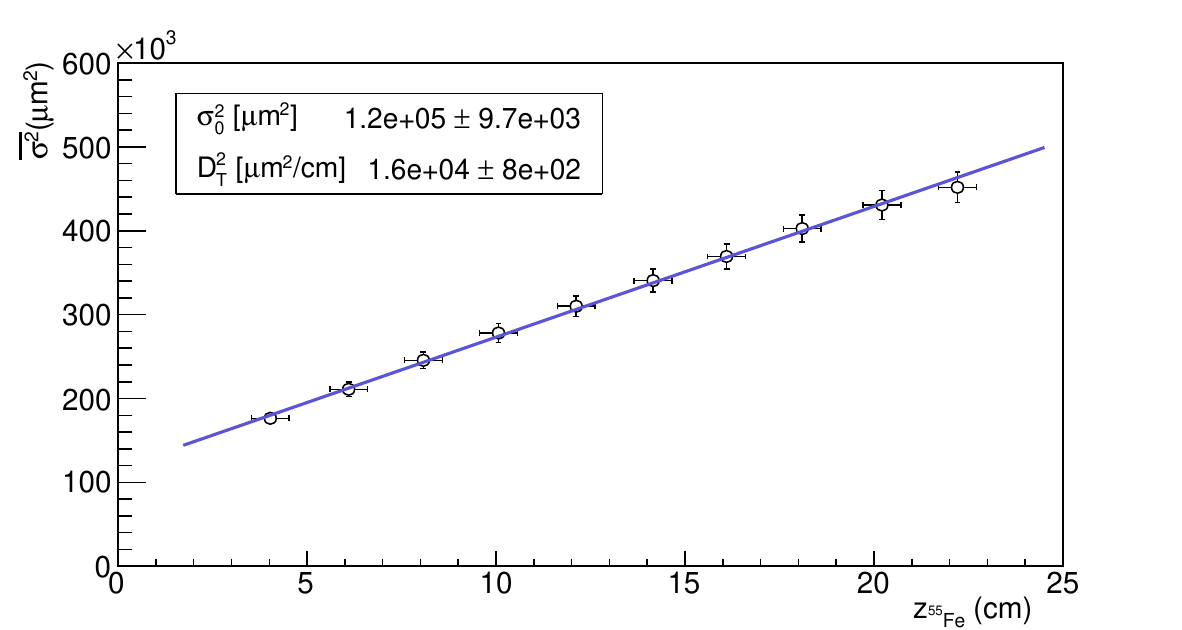}
    \caption{Average values of the $\sigma$ distribution, squared, as a function of the  position \zfe, with a superimposed linear fit. Vertical uncertainties on the points are the statistical uncertainties from the Gaussian fits, while horizontal ones describe the spread in $z$ of the interactions as described in Sec. \ref{sec:pos}.}
    \label{fig:sigma}
\end{figure}

The mean values of the $\sigma$ distributions, squared, are shown  in the plots in Fig.~\ref{fig:sigma} as a function of the distance $z$ of the source from the GEMs.

A linear fit is superimposed, yielding the following estimates for \(\sigma_0\) and the transverse diffusion coefficient (D$_{T}$) at \SI{1.0}{kV/cm}:
\begin{equation}
    \sigma_0 = (340 \pm 14) \, \mu \mathrm{m}
\end{equation}
\begin{equation}
    D_T = (125 \pm 5) \, \frac{\mu \mathrm{m}}{\sqrt{\mathrm{cm}}}
\end{equation}

These values  are in  good agreement with what was evaluated with Garfield ($D_T = (116) \, \frac{\mu \mathrm{m}}{\sqrt{\mathrm{cm}}}$) and measured ($\sigma_0 = (280 \pm 60) \, \mu \mathrm{m}$) with a different setup by our collaboration \cite{bib:stab}.

\subsection{Spot light}

\begin{figure}[h!]
    \centering
    \includegraphics[width=0.49\linewidth]{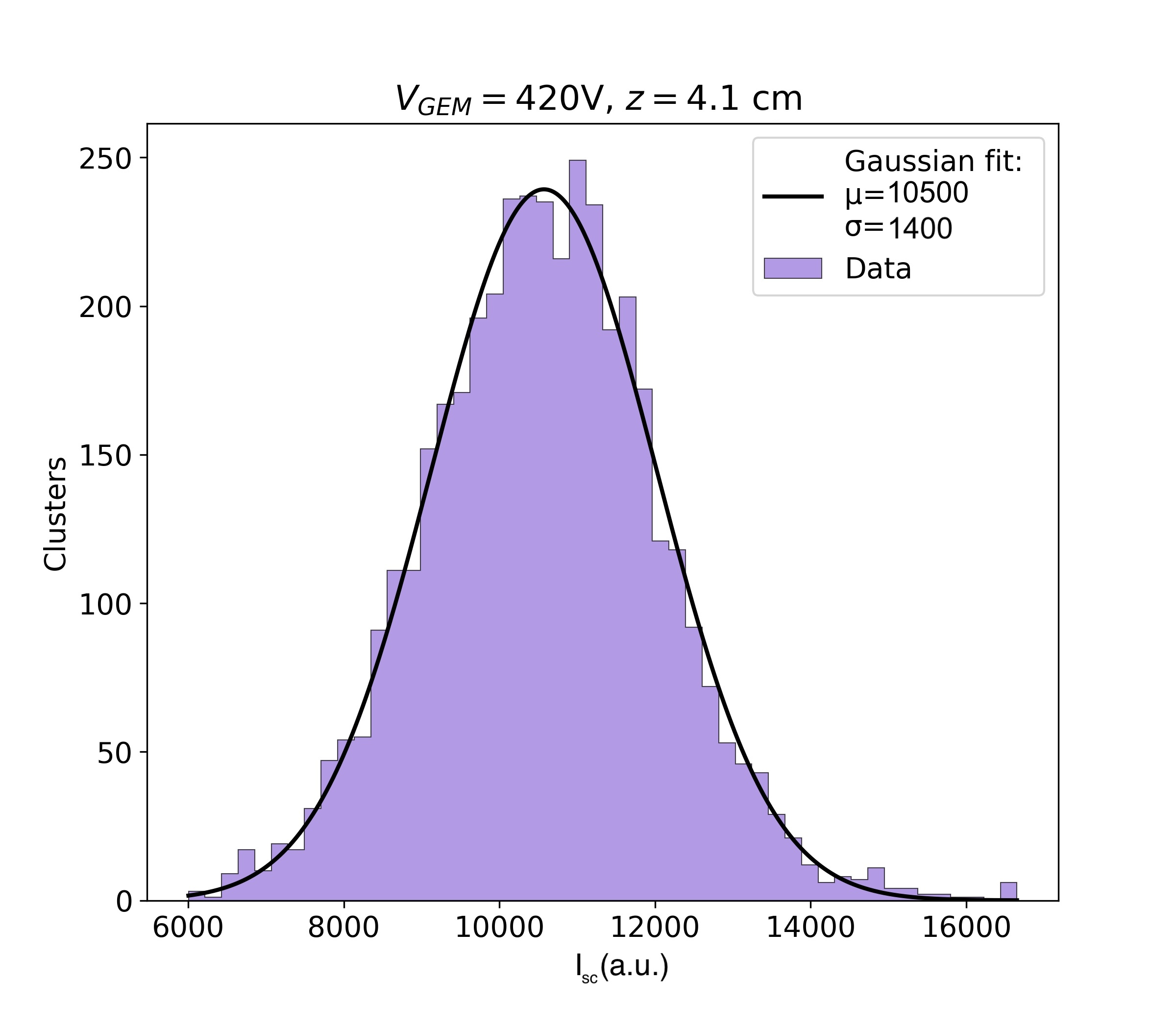}
    \includegraphics[width=0.49\linewidth]{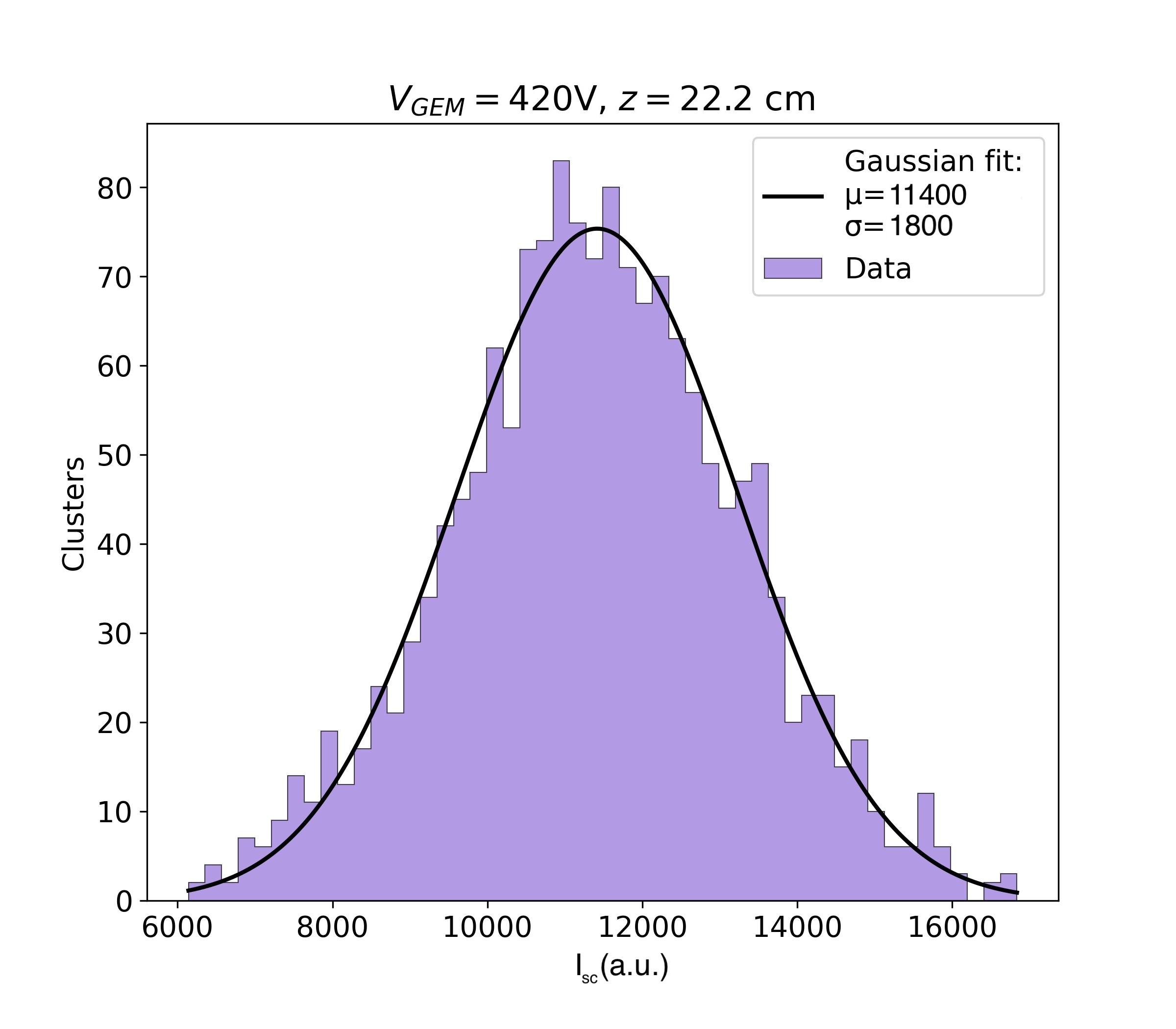}     \\
    \includegraphics[width=0.49\linewidth]{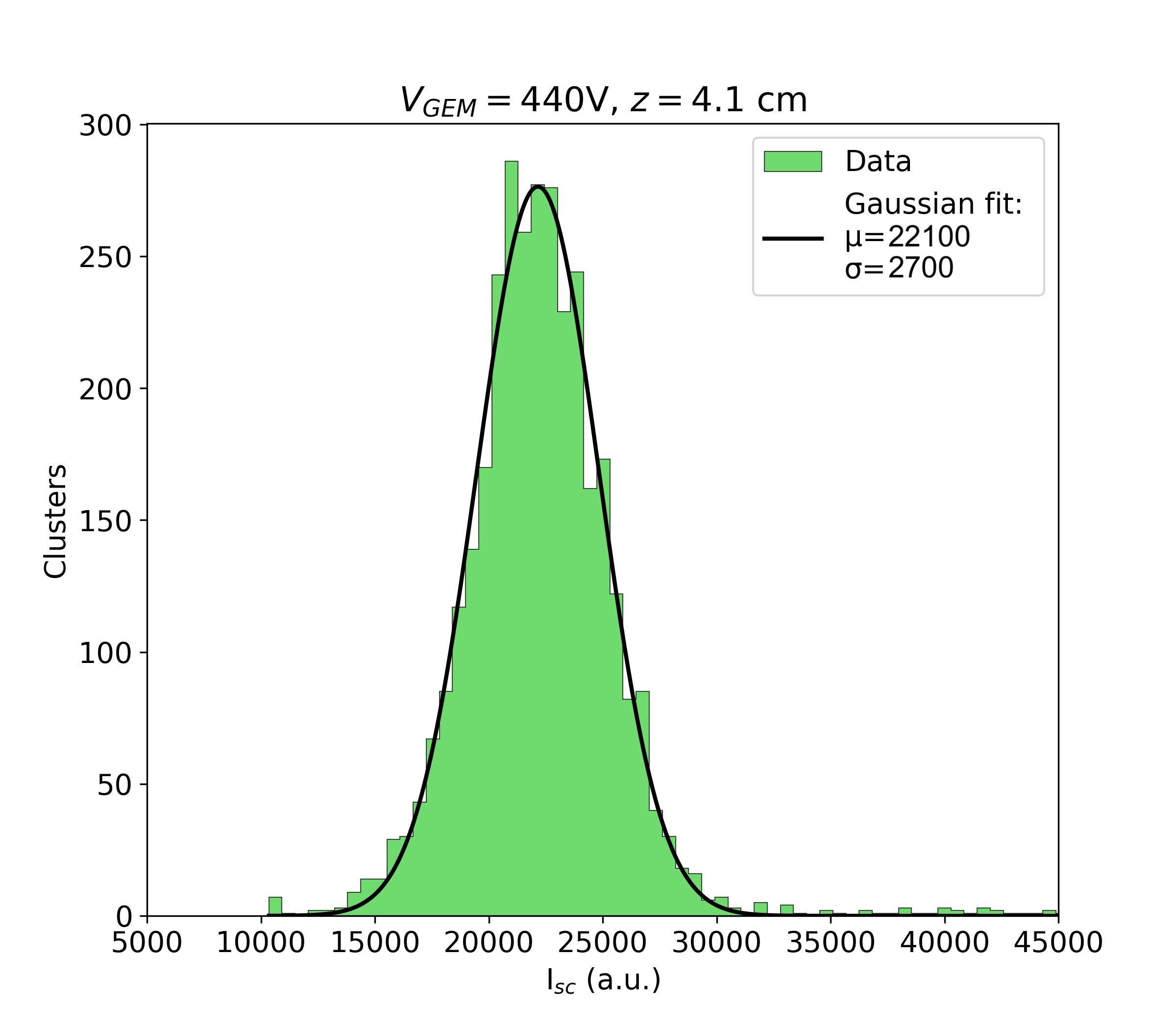}
    \includegraphics[width=0.49\linewidth]{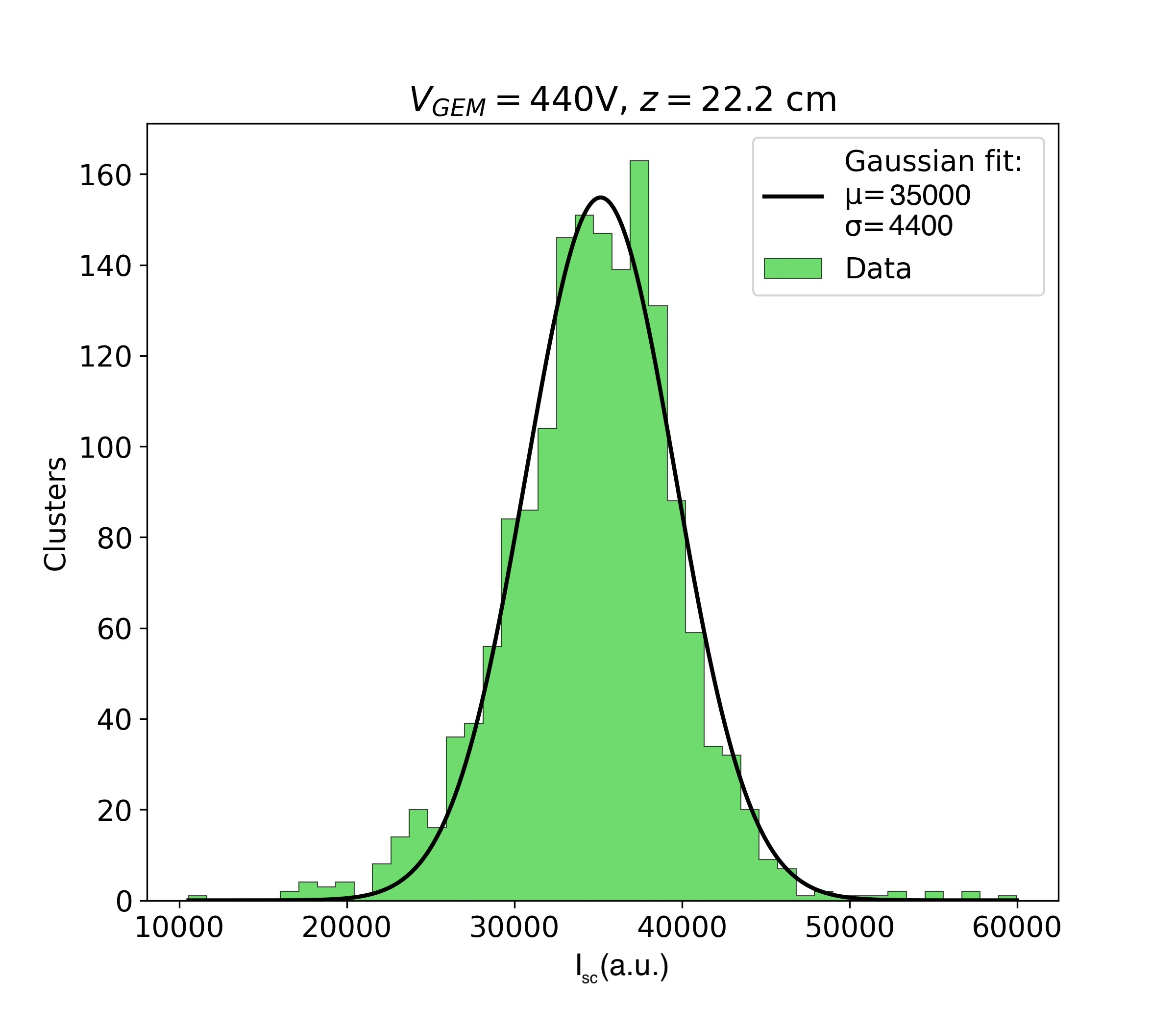} 
    \caption{Distributions of the values of the light integral (\Isc) evaluated for all clusters attributed to the \fe interactions for \Vg=\SI{420}{V} (top) and \Vg=\SI{440}{V} (bottom) with the source in two different positions with a gaussian fit superimposed.}
    \label{fig:hint}
\end{figure}

The distributions of \Isc values, representing the total light in the spot, were also studied for different values of \zfe and \Vg (making all GEMs working at the same voltage).
In particular for \Vg~=~\SI{440}{V}, data were taken for the same source positions described in the previous section, while other data at a voltage of \Vg~=~\SI{430}{V} and \Vg~=~\SI{420}{V}, were only taken for \zfe~=~\SI{4}{cm}, \SI{8}{cm}, \SI{12}{cm}, \SI{16}{cm}, and \SI{20}{cm}.
For each run, the distribution of  \Isc for all the clusters satisfying the \fe selection is filled, and is shown in Fig.~\ref{fig:hint} for different \Vg and \zfe.

As described in Sec.~\ref{sec:optical} and in Eq.~\ref{eq:fiorina1}, a direct proportionality between the light in the spot and the number of primary ionised electrons is expected. 
Therefore, while different values of \Vg are expected to have a large impact on the total amount of charge and light produced, since the gain should be independent from the event depth in the gas volume, the mean values of the aforementioned distributions should ideally remain independent from \zfe. 

From Fig. \ref{fig:hint} it is possible to see how, instead, the amount of collected light increases as \zfe increases for all values of \Vg.

\begin{figure}[h!]
    \centering
    \includegraphics[width=1.1\linewidth]{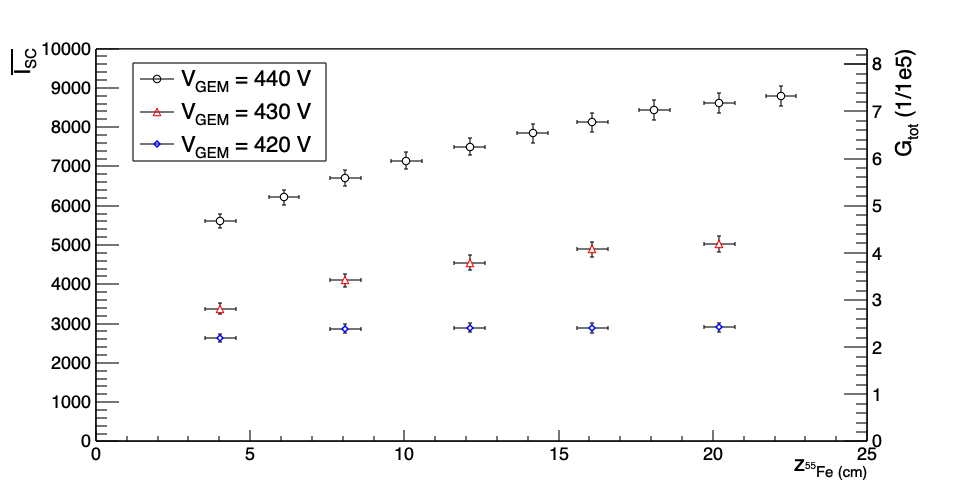}
    \caption{Average \Isc and gas gain values measured at different distances from the GEM plane for \Vg values of 440 V (black), 430 V (red), and 420 V (blue). The horizontal bars indicate the spread in $z$ of the photon interaction points as described in Sec. \ref{sec:pos}; the vertical bars take into account not only the statistical fluctuations of the mean values of the gaussian fits, but also other  uncertainties (due for example to environmental conditions) and evaluated by comparing independent measures repeated in the same configurations.}
    \label{fig:cam_satu}
\end{figure}

The behaviour of \Isc and the corresponding  G$_{tot}$ values evaluated  by means of Eq.~\ref{eq:fiorina1}, for different \zfe and three different \Vg values (all \Vg = \SI{440}{V}, \SI{430}{V} and \SI{420}{V}) are shown in Fig.~\ref{fig:cam_satu} as a function of the source position. 
Vertical bars account for the statistical fluctuations of the mean values of the gaussian fits and for other uncertainties (due for example to environmental conditions) that were evaluated by comparing independent measures repeated in the same configurations.

On the one hand, it is clearly observed that the total light and the gain, for each value of \zfe, depend on \Vg and increase with it. On the other hand, it is also noticeable that the detector response is not constant and tends to increase for high values of \zfe in all cases.
This effect was already observed and described in other papers by the CYGNO collaboration \cite{bib:lime_over}. A similar effect had already been observed by other research groups \cite{bib:saturationCERNMIC,bib:satuRoy}.
As described in the next section, our interpretation is that the observed behaviour can be explained as a \textit{gain saturation} occurring for GEMs operating in very high gain conditions and producing a high charge density in the multiplication channels. The electron diffusion, by reducing the charge spatial density, can help in mitigating this effect leading to an increase in the actual gain for ionizations occurring farther from the GEM plane.

\section{Gain saturation model}

A simple model was developed by the \cygno collaboration to microscopically explain the phenomena leading to this gain saturation.

Let us assume that, during the development of the avalanche within the GEM multiplication channels, a significant amount of electrons and positive ions are produced. Under the effect of the electric field present in the channel, these last slowly migrate towards the lower potential plane of the GEM. 
Their total positive electric charge $Q_p = nq$ (where $n$ is the amount of ion-electron pairs produced and $q$ is the charge of the single ions)  will partially shield the electric field present in the channel.
This effect was already simulated with a Finite Element Analysis program \cite{bib:saturationCERNMIC}. From that study authors concluded that the presence of ions reduces the amplification field affecting the multiplication processes.
This reduction of the effective field in the GEM channel can directly reduce the avalanche development or favor a recombination phenomenon. Even if from the data it is not possible to distinguish between these two cases, the final effect is that the actual gain of each channel will  not only depend on its operation voltage \Vg, but also on how much charge is produced.
At first order, we can assume that a uniform electric field is present within the GEM channels  \Eg = \Vg$/d$ where $d$ is the GEM thickness~\cite{bib:thesis}.
Electrons moving toward the higher potential plane of the GEM will then feel an effective electric field $E$ due to the superposition of \Eg and the one produced by positive charge expected to be proportional to $Q_p$: $E_p = \gamma Q_p$.
Therefore: 
\begin{equation}
E=E_{\mathrm{GEM}}-E_p= E_{\mathrm{GEM}}\cdot\left(1-\frac{E_pd}{V_{\mathrm{GEM}}}\right)
\end{equation}
We can introduce a screen parameter of the field, denoted as $\beta = r/$\Vg (with $r=\gamma q d)$, so that
at any given time, the electric field effectively accelerating the charges in the channel is given by \Eg~$\cdot~(1-\beta n)$. 

The increase of the number of charges $n$ in a spatial step $ds$ is usually described by Townsend  equation \cite{bib:Leo}:
\begin{equation}
    \frac{dn}{ds} = \alpha  n
\end{equation}
where $\alpha$ represents the inverse of the mean free path for the ionization process and has a strong dependence on the electric field. At first order, this dependence can be assumed as linear \cite{bib:2001mug,bib:diethorn}: 

\begin{equation}
\alpha(E) = \tilde{\alpha} \cdot (E - E_{0})
\end{equation}
where $\tilde{\alpha}$ has the dimension of the inverse of an electrical potential and $E_{0}$ represents the electric field value around which linear expansion is made.
The Townsend equation, in presence of the screening effect of the ions during the multiplication process, would then become:
\begin{equation}
\frac{dn}{ds} = \tilde{\alpha} \cdot (E_{\mathrm{GEM}}(1-\beta n) - E_{0}) n
\end{equation}
By integrating this equation, inside the GEM channel:
\begin{equation}
    \int^{n_{out}}_{n_{in}} \frac{dn}{(E_{\mathrm{GEM}}(1-\beta n) - E_{0}) n} = \int^d_0 \tilde{\alpha}  ds
\end{equation}
we can obtain the formula to evaluate the gain of a GEM channel $G=n_{out}/n_{in}$ with a voltage drop \Vg :

\begin{equation}
\label{eq:gain}
        G = \frac{e^{\tilde{\alpha} (V_{\mathrm{GEM}}-V_0)}}{1 + \beta' n_{in} (e^{\tilde{\alpha} (V_{\mathrm{GEM}}-V_0)} - 1)}.
\end{equation}
where:

\begin{itemize}
    \item  $V_{0}$ is equal to E$_{0}/d$;
    \item a new parameter $\beta'$ is introduced $\beta' = \frac{\beta V_{\mathrm{GEM}}}{V_{\mathrm{GEM}} - V_{0}} = \frac{r}{V_{\mathrm{GEM}} - V_{0}}$  which shows that the relevant electrical potential is the difference between \Vg and $V_{0}$.
\end{itemize}

The \textit{no-saturated gain} showing the usual exponential behaviour foreseen by the Townsend formula can be indicated as $\tilde{G} = e^{\tilde{\alpha} (V_{\mathrm{GEM}}-V_0)}$.

The equation \ref{eq:gain} can be rewritten as:
\begin{equation}
\label{eq:gain0}
        G = \frac{\tilde{G}}{1 + \beta' n_{in} (\tilde{G} - 1)}
\end{equation}

It should be noticed that $G$ depends on the product of the parameter $\beta'$ and the number $n_{in}$ of primary electrons entering the channel  and on \Vg. 
In particular, it can be observed that:

\begin{itemize}
    \item \textbf{(a)} if $\beta' n_{in} \simeq 0$ (i.e. negligible screen effect), G is equal to $\tilde{G}$ (the no-saturated gain);
    \item \textbf{(b)} if $\beta' n_{in} \simeq 1$ then $G \simeq 1$, the saturation fully suppresses the gain in the GEM channel, giving, ideally, $n_{out}=n_{in}$.
\end{itemize}


Let us now assume that in the first two GEMs (GEM$_1$ and GEM$_2$), the total amount of charge is low enough to be in the case (a) above (i.e. $G_1 = \tilde{G}_1 $ and $G_2=\tilde{G}_2$) and that the gain saturation only affects GEM$_3$. 

Electrons ionised in the gas volume at a certain $z$, after the drift and the multiplication processes in the first two GEMs, produce an electron cloud with a space distribution that can be modeled as a 3D normal distribution with variance $\sigma$. 
The volume of the  electron cloud  will then be  proportional to $\sigma^3$. 
In general,  efficiency in collecting ($\epsilon^{coll}$) and extracting ($\epsilon^{extr}$) electrons in the GEM channels is less than one \cite{bib:bachmann,bib:bonivento}. For each GEM$_i$ we can indicate with $\epsilon_i = \epsilon^{coll}_i \cdot \epsilon^{extr}_i$ the product of these efficiencies. Since electrons are not extracted from GEM$_3$, the relevant parameter in that case is $\epsilon_3 = \epsilon^{coll}_3$.

Thus, in GEM$_3$, the amount of charge collected by each channel $n_{in}$:
\begin{itemize}
\item depends on the primary ionisation in the gas $n_e$; 
\item is proportional to  $k/\sigma^3$ where $k$ takes into account the GEM channels dimensions and density;
\item increases as the product of the gains and of the efficiencies $\epsilon_1 G_1 \epsilon_2 G_2\epsilon_3$;
\end{itemize}

In the case that all GEMs operate at the same \Vg, $\tilde{G}_1$ = $\tilde{G}_2$ = $\tilde{G}_3$ = $c^{\tilde{\alpha} (V_{\mathrm{GEM}}-V_0)}$ that we can indicate as $G$.
The total gain of the triple-GEM stack can be written as:
\begin{eqnarray}
\label{eq:gain-fit}
 G_{tot} = \epsilon_1G_1\cdot \epsilon_2G_2 \cdot \epsilon_3G_3 = ~~~~~~~~~~~~~~~~~~~~~~~~~~~~~~~~~\\
 = \epsilon\cdot \tilde{G}_1 \cdot \tilde{G}_2 \cdot  G_3 = ~~~~~~~~~~~~~~~~~~~~~~~~~~~~~~~~~~~~~~~\\
 = \frac{\epsilon \tilde{G}^3}{1+\epsilon p\cdot \tilde{G}^2(\tilde{G}-1)/(\sigma^3(V_{\mathrm{GEM}}-V_{0}))} ~~~~~~\\  
= \frac{\tilde{G}^3\sigma^3}{\sigma^3/\epsilon+p\cdot\tilde{G}^2(\tilde{G}-1)/(V_{\mathrm{GEM}}-V_{0})}   ~~~~~~
\end{eqnarray}
%
%
where the  parameter $p = rk$ and $\epsilon = \epsilon_1\cdot \epsilon_2 \cdot \epsilon_3$

\begin{figure}[h!]
    \centering
    \includegraphics[width=0.99\linewidth]{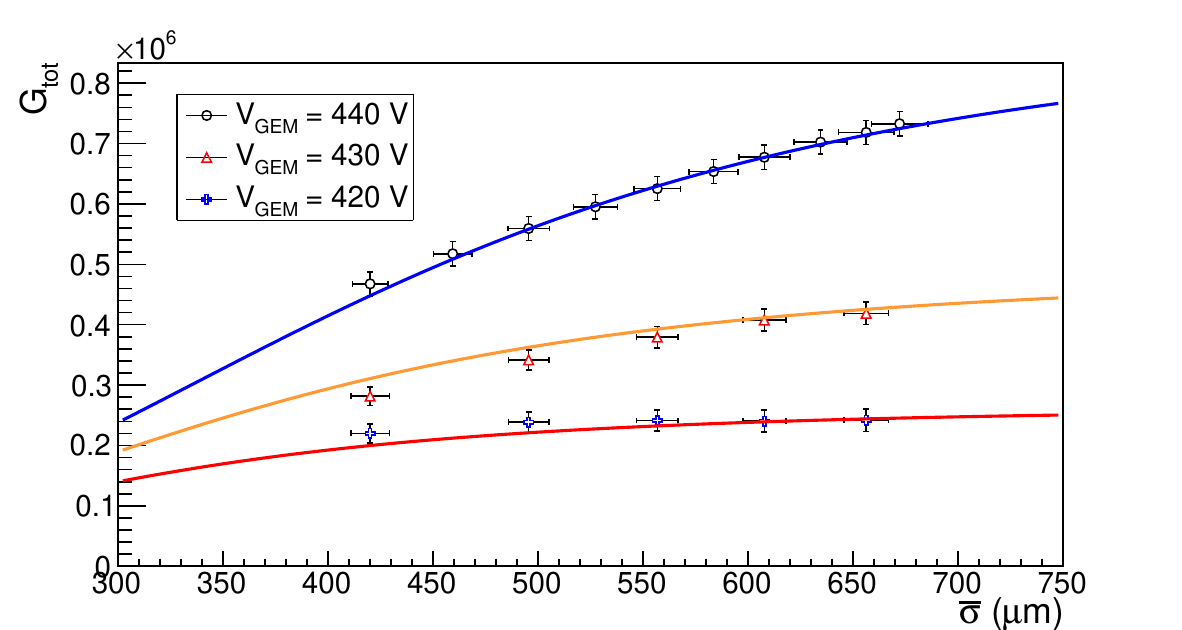}
    \caption{Charge gain as a function of the average spot $\sigma$ obtained from the fit to the distributions measured on \fe deposits. Values are shown for the Triple-GEM stack in three configurations: all \Vg~=~\SI{440}{V}, all \Vg~=~\SI{430}{V} and all \Vg~=~\SI{420}{V} with superimposed curves representing the results of a \textit{simultaneous} fit to the three parameters entering Eq.~\ref{eq:gain-fit}.}
    \label{cam_satu-fit}
\end{figure}

Figure~\ref{cam_satu-fit} illustrates the behaviour of $G_{tot}$ as a function of $\bar{\sigma}$, as measured in the data for all GEMs at \Vg = \SI{440}{V}, \Vg = \SI{430}{V}, and \Vg = \SI{420}{V}. These measurements are  the same entering in Figs.~\ref{fig:sigma} and \ref{fig:cam_satu}. 

The curves overlaid on the data points represent the results of a  fit performed with MINUIT simultaneously on the three datasets at different \Vg in order to obtain a single set of the four parameters  entering Eq.~\ref{eq:gain-fit} ($\alpha$, $V_0$, $p$ and $\epsilon$) that could describe the experimental data.

The fit procedure returns a $\chi^2$ value normalised to the number of the degrees of freedom of 0.7 with a pValue of 0.81.

The values of the parameters calculated from the fit are described below: 

\vspace{10pt}
$\tilde{\alpha} = (2.05 \pm 0.02) \times 10^{-2}\SI{}{V^{-1}}$,

$V_{0} = 200 \pm$~\SI{10}{V}, 

$p = (2.0 \pm 0.2) \times 10^{4} ~\SI{}{\mu m^3 V}$,

$\epsilon = 0.38 \pm 0.06$
\vspace{10pt}

The values of $\alpha$ and $V_{0}$ obtained from this study are in reasonable agreement with results found in experimental studies by the \cygno collaboration in different setups (e.g. in \cite{bib:roby} with a measurement performed by monitoring the detector currents under the irradiation of an X-ray tube, an $\tilde{\alpha}$ = $2.3 \pm 0.1$ \SI{}{V^{-1}} and  a $V_{0}$ = $160 \pm 15$ \SI{}{V} were obtained).

These parameters also allow for the evaluation of the  no-saturated gains ($\tilde{G}$) for the tested \Vg values:

\vspace{10pt}
$\tilde{G}(\mathrm{V}{_\mathrm{GEM}}= 440{\mathrm{V}}) =135~\pm~12$

$\tilde{G}(\mathrm{V}{_\mathrm{GEM}}= 430{\mathrm{V}}) =105~\pm~10$

$\tilde{G}(\mathrm{V}{_\mathrm{GEM}}= 420{\mathrm{V}}) =90~\pm~8$
\vspace{10pt}

As discussed in the introduction, the main purpose of this work is the need to predict the gain of the detector as a function of charge density, in order to simulate the detector response for different released energies and different drift distances.
To assess its reliability on this goal, the distribution of the fit residuals to the data, normalised by the data values, was studied. This distribution shows an RMS value of 0.04 indicating  that, across the entire range of gain values, spanning approximately from $2 \times 10^{5}$ to $7 \times 10^{5}$, the proposed model is able to foresee the behavior of gain values as a function of the charge density with a relative precision of 4\%.

Considering the approximations made (namely, that the electric field in the channel is uniform and that the charge distribution within the cloud is uniform, resulting in a constant density), this result indicates that the proposed model 
 can represent a promising tool for numerical simulations of the gain  to predict its behaviour under conditions of gain and release of energies also different from those of this work.



\section{Conclusion}

In this study, the light response of an optically read-out GEM-based TPC developed for the CYGNO experiment  was investigated. Through a systematic analysis of experimental data collected with a two-liter prototype, the dependence of the response of detector on the charge density and spatial distribution of ionization electrons was studied and characterised.

From the study of the shape and the total amount of photons of the light spots induced by \fe X-rays,  key parameters such as transverse diffusion and light yield variations as a function of the drift distance were extracted. 
 A phenomenological model describing gain saturation effects due to space-charge accumulation was developed, which successfully reproduces the observed trends in experimental data with a precision of a few percent.

These findings validate the proposed mathematical model as a reliable tool capable of predicting the response behavior that can be used to simulate the response of GEM detectors in presence of high primary charge density.

The observations described in this paper confirm a gain-reduction phenomenon inside the GEM channels, which can be attributed to the presence of a high charge density. This charge modifies the local electric field, which can, in principle, lead to a reduced avalanche development efficiency or to recombination phenomena.

The results obtained in this study can be considered a starting point for studying the phenomenon in greater detail and attempting to identify and quantify the possible presence of charge recombination processes, and also for suggesting possible solutions that would make the response linear even at high ionization densities, contributing to the development and optimisation of the design of optically read-out TPC.

\begin{acknowledgement}
\textbf{Acknowledgement}

This project has received fundings under the European Union’s Horizon 2020 research and innovation program from the European Research Council (ERC) grant agreement No 818744 and is supported by the Italian Ministry of Education, University and Research through the project PRIN: Progetti di Ricerca di Rilevante Interesse Nazionale “Zero Radioactivity in Future experiment” (Prot. 2017T54J9J).
We want to thank General Services and Mechanical Workshops of Laboratori Nazionali di Frascati (LNF).
\end{acknowledgement}

 
\bibliography{mybiblio}

\end{document}